\documentclass[prl,aps,showpacs,twocolumn,floatfix]{revtex4-1}
\usepackage{amsmath, amsfonts, amssymb, bm, graphicx, color, mathtools}

\newcommand{\bS}{{\bm S}}

\newcommand{\bsig}{{\bm \sigma}}

\newcommand{\e}{{\rm e}}
\newcommand{\ii}{{\rm i}}
\newcommand{\dd}{d^\dag}

\newcommand{\cd}{c^\dag}

\newcommand{\ua}{{\uparrow}}
\newcommand{\da}{{\downarrow}}
\newcommand{\vac}{\vert {\rm vac} \rangle}

\begin{document}

\title{Spin-orbit coupled correlated metal phase in Kondo lattices: an
       implementation with alkaline-earth atoms}

\author{L. Isaev}
\author{J. Schachenmayer}
\author{A. M. Rey}
\affiliation{JILA, NIST, Department of Physics \& Center for Theory of Quantum
Matter, University of Colorado, 440 UCB, Boulder, CO 80309, USA}

\begin{abstract}

 We show that an interplay between quantum effects, strong on-site
 ferromagnetic exchange interaction and antiferromagnetic correlations in Kondo
 lattices can give rise to an exotic spin-orbit coupled metallic state in
 regimes where classical treatments predict a trivial insulating behavior.
 This phenomenon can be simulated with ultracold alkaline-earth fermionic atoms
 subject to a laser-induced magnetic field by observing dynamics of spin-charge
 excitations in quench experiments.

\end{abstract}

\maketitle

\paragraph*{Introduction.--}

The behavior of correlated quantum systems can rarely be understood in terms of
individual atoms or electrons, and instead is determined by a competition
between their strong interactions and kinetic energy \cite{anderson-1972-1}.
This interplay often places states with fundamentally different properties
energetically close to each other, and hence makes the system highly sensitive
to external controls such as pressure, or magnetic field \cite{morosan-2012-1}.

A paramount example of correlation-driven tunable phenomena is the colossal
magneto-resistance in transition-metal oxides, e.g. manganites
\cite{dagotto-2005-1,dagotto-2005-2}.
Properties of these materials are governed by the ferromagnetic Kondo lattice
model (FKLM) which includes competition between kinetic energy of itinerant
electrons and their Hund exchange coupling with localized spins
\cite{izyumov-2001-1,edwards-2010-1}.
This interaction often exceeds the conduction bandwidth and ensures that only
on-site triplets, i.e. electrons whose spins are aligned with local magnetic
moments, can exist at low energy.
For classical core spins ($S \gg 1$), an effective electron hopping amplitude
between two lattice sites strongly depends on the magnetic background: it is
largest when local spins on the two sites are parallel, and vanishes for
anti-parallel [antiferromagnetically (AF) ordered] local moments
\cite{anderson-1955-1,degennes-1960-1}.
As a result, the conductivity of a system becomes highly sensitive to small
variations in the magnetic texture, e.g. caused by an external magnetic field.
This so-called double-exchange (DE) physical picture remains qualitatively
valid when quantum fluctuations of the local magnetism are taken into account
\cite{dagotto-1996-1,zang-1997-1,brunton-1998-1,dagotto-1998-1,shannon-2002-1}
and in the extreme quantum case $S = \frac{1}{2}$ \cite{kienert-2006-1}.

Nevertheless, even early works \cite{degennes-1960-1} hinted at a breakdown of
the DE semiclassical description in the presence of strong AF correlations
between local spins when they form at least short-range N\'eel order.
While in an ideal antiferromagnet an electron can not move, it still gains
energy via smooth deformations of the N\'eel background \cite{degennes-1960-1}.
Quantum fluctuations would allow local spins to form singlets with mobile
fermions and further distort the AF texture.
Even for Hund coupling comparable to the conduction bandwidth these processes
are important and can lead to an increase of the electron effective mass
\cite{lehur-2007-1}.

\begin{figure}[t]
 \begin{center}
  \includegraphics[width = \columnwidth]{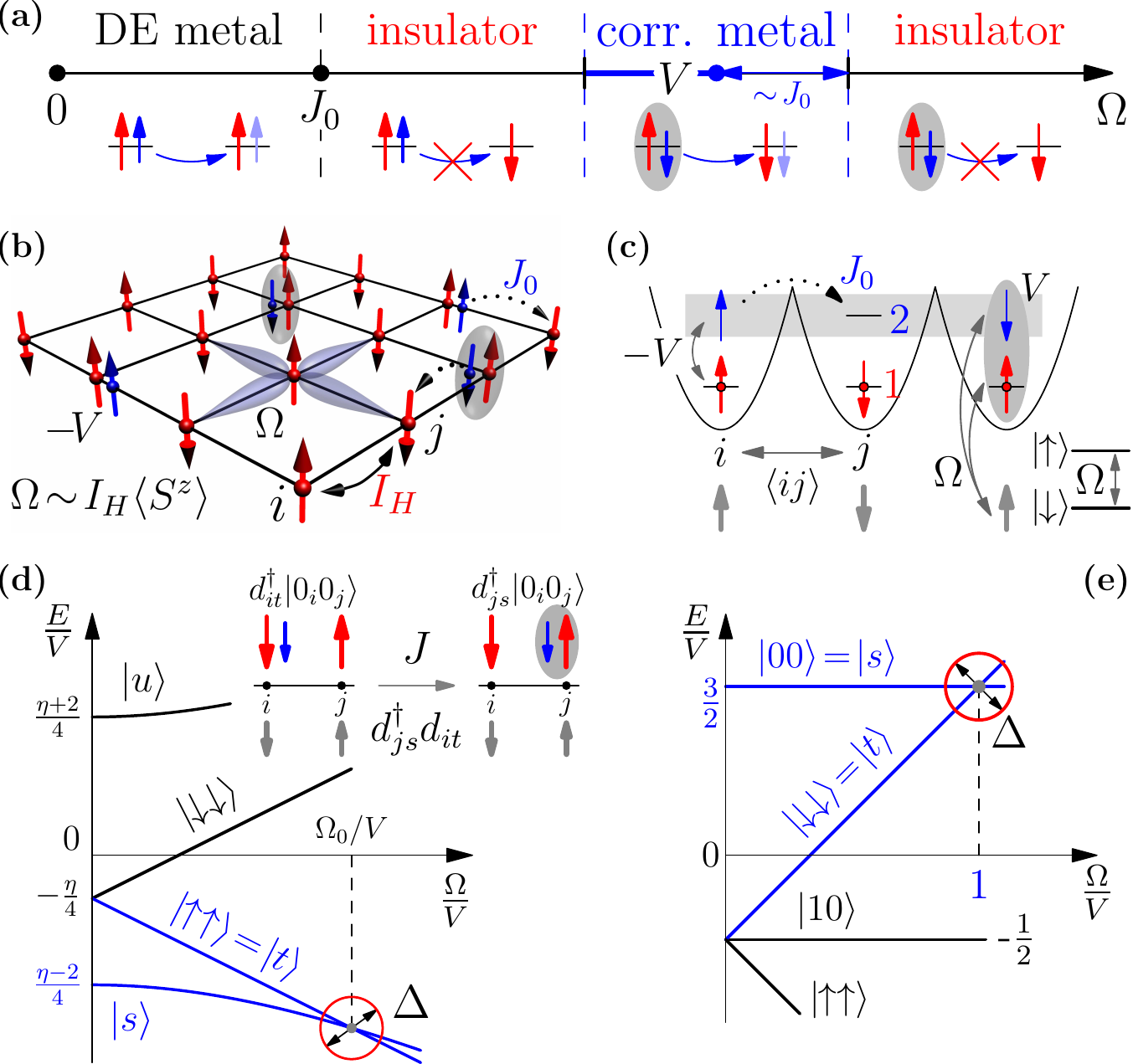}
 \end{center}
 \caption{
  {\bf (a)} Schematic phase diagram of Eq. \eqref{eq:fm-xxz-klm} with $V \gg
  J_0$ ($\Omega$ is the AF interaction strength), featuring the correlated
  metal state for strong coupling $\Omega \sim V$.
  The conventional ferromagnetic (DE) metal occurs at $\Omega \lesssim J_0$.
  Bottom row: allowed and forbidden (indicated by red crosses) hopping
  processes.
  Blue (red) color marks mobile (local) fermions.
  Light blue spins show final states of itinerant fermions.
  Gray ellipses denote local entangled singlet-triplet (s-t) states [$\vert s
  \rangle$ in (d)].
  {\bf (b)} The electronic model of Eq. \eqref{eq:fm-xxz-klm}.
  Color notations are as in (a).
  Local spins feel a staggered mean-field $\Omega \sim I_H \langle S^z \rangle$
  due to the AF background.
  {\bf (c)} Optical lattice of the AEA setup Eq. \eqref{eq:klm}.
  Band 1 (2) is localized (itinerant).
  Gray arrows indicate the laser-induced staggered Zeeman field $\Omega$ that
  splits pseudo-spin $\vert \ua, \da \rangle$ states.
  Gray ellipse is a spin-singlet state.
  Other notations are as in (b).
  {\bf (d)} Energies on an isolated site with one fermion.
  Blue lines [red circle] indicate the s-t subspace [resonance].
  Inset: Hopping of s-t excitations $d_{i \alpha}$ \eqref{eq:d-fermions}.
  Other notations are as in (a).
  {\bf (e)} Same as in (d), but for the AEA model \eqref{eq:klm}.
  The s-t subspace is an excited manifold.
 }
 \label{fig:fig1}
\end{figure}

In the present Letter we demonstrate that quantum nature of the local magnetism
{\it dramatically affects} physics of a FM Kondo lattice with AF correlations
between core spins [Fig. \ref{fig:fig1}(a)].
We focus on a $S = \frac{1}{2}$ FKLM in the strong-coupling regime, when Hund
and AF interactions exceed the electron bandwidth, and show that the AF
environment of each core moment frustrates the on-site Hund exchange $V$ [Fig.
\ref{fig:fig1}(b)].
Properties of the model are controlled by a competition between $V$ and an
energy scale $\Omega$ of the antiferromagnetism.
When these energies are significantly different, the system is an insulator
with localized band electrons.
However, near the resonance $V \approx \Omega$, the AF and Hund
interactions effectively cancel each other allowing quantum effects to
stabilize a new {\it correlated metal} phase whose quasiparticles admix singlet
and triplets states of bare electrons and local spins.
These excitations distort the AF order and give rise to a {\it transverse} (to
the N\'eel vector) magnetization.
This resonant behavior is absent in a semiclassical DE theory which predicts an
insulating state for any Hund coupling.

The correlated metal phase can be observed in fermionic alkaline-earth atoms
(AEAs) \cite{gorshkov-2010-1}, such as ${}^{87}{\rm Sr}$ \cite{zhang-2014-1} or
${}^{173}{\rm Yb}$ \cite{cappellini-2014-1,scazza-2014-1,pagano-2015-1,
hofer-2015-1,zhang-2015-1,zhang-2015-2}, in a two-band optical lattice where
atoms in the lowest (localized) and higher (itinerant) bands correspond to core
spins and mobile fermions, respectively [Fig. \ref{fig:fig1}(c)].
We propose to simulate the AF background with an artificial, laser-induced
magnetic field \cite{dalibard-2011-1,celi-2014-1,mancini-2015-1}, which in AEAs
can be implemented either using Raman transitions between nuclear spin levels
\cite{cooper-2015-1} or Rabi coupling of ${}^1 S_0$ and ${}^3 P_0$ electronic
clock states \cite{wall-2016-1}.
The laser phase can be controlled to vary from one lattice site to the next in
a {\it staggered} fashion, while the Rabi or Raman coupling of relevant atomic
states provides a handle of the above singlet-triplet (s-t) resonance.

\paragraph*{Correlated metal in a strongly-coupled FKLM.--}

Let us consider a generic FM Kondo lattice with AF superexchange interactions
between core moments:
\begin{align}
 H = & -J_0 \sideset{}{_{\langle i j \rangle}}\sum (\cd_{i n} c_{j n} +
 {\rm h.c.}) + I_H \sideset{}{_{\langle i j \rangle}}\sum S^z_i S^z_j -
 \nonumber \\
 & - V \sideset{}{_i}\sum \bigl[ \bS^\perp_i \cdot {\bm s}^\perp_{c i} +
 \eta S^z_i s^z_{c i} \bigr] + R \sideset{}{_{\langle i j \rangle}} \sum n^c_i
 n^c_j,
 \label{eq:fm-xxz-klm}
\end{align}
defined on a bipartite (e.g. square) lattice of Fig. \ref{fig:fig1}(b).
Here $\cd_{i n}$ creates an electron with spin $n = \ua$, $\da$ (we assume
summation over repeated indices) at site $i = 1 \ldots N$.
The first line contains nearest-neighbor (NN) hopping $J_0$ on a link $\langle
i j \rangle$, and the AF exchange $I_H > 0$ between local moments $\bS_i$.
The latter are coupled to mobile spins ${\bm s}_{c i} = \frac{1}{2}
\cd_{i n} \bsig_{n m} c_{i m}$ ($\bsig$ are Pauli matrices) via a FM Hund
exchange $V$ with an $XXZ$ anisotropy ($\perp$ denotes $x y$ vector components)
$\eta \in [0, 1]$ arising from atomic spin-orbit coupling and crystal-field
effects.
Due to same reasons, the AF interaction is also anisotropic: we consider the
simplest Ising limit, but our results are applicable to a general $XXZ$ case.
The last term (with $n^c_i = \cd_{i n} c_{i n}$) is a NN Coulomb repulsion (see
below).

We focus on the strong-coupling limit $J_0 \ll I_H$, $V$ and assume that core
moments are AF-ordered, $\langle S^z_j \rangle = \langle S^z \rangle
\e^{\ii {\bm Q} \cdot {\bm x}_j}$ [${\bm x}_j \equiv j$, ${\bm Q} = \pi$ for
a one- (1D), $(\pi, \pi)$ for a two-dimensional (2D) lattice, etc].
Mobile electrons will form entangled states with local spins, and above certain
density $n^c > n^c_{cr}$, destroy the N\'eel order even for $J_0 \equiv 0$.
Hence, the above assumption is valid only in the low-density regime $n^c \ll 1$
when electrons rarely occupy NN sites.
This regime is enforced by the repulsion $R$ in Eq. \eqref{eq:fm-xxz-klm}.
The AF background can be taken into account by performing a staggered
transformation:
\begin{align}
 c_{i n} = & a_{i n} \,\,\, (i \,\, {\rm even}); \quad
 c_{i n} = a_{i, -n} \,\,\, (i \,\, {\rm odd}); \nonumber \\
 \bS_i = & \bigl( T_i^x, (-1)^i T^y_i, (-1)^i T^z_i \bigr),
 \label{eq:st-trans}
\end{align}
where $a_{i n}$ and ${\bm T}_i$ are new mobile fermions and local spins. As a
result, the first line in Eq. \eqref{eq:fm-xxz-klm} becomes
$-\sum_{\langle i j \rangle} \bigl[ J_0 \sigma^x_{n m} (a^\dag_{i n} a_{j m} +
{\rm h.c.}) + I_H T^z_i T^z_j\bigr]$; other terms remain unchanged with
$c_{i n} \to a_{i n}$ (${\bm s}_{c i} \to {\bm s}_{a i}$) and
$\bS_i \to {\bm T}_i$.

In this staggered frame, the mean-field Hamiltonian of an isolated site is
$H_i = -V \bigl[ {\bm T}^\perp_i \cdot {\bm s}^\perp_{a i} + \eta T^z_i
s^z_{a i}\bigr] - z I_H \langle T^z_j \rangle T^z_i$.
Here the last term is a molecular field acting on a core spin in the AF
environment and $z$ is the lattice coordination number.
We assume that $\langle T_j^z \rangle = \langle T^z \rangle > 0$ is
$j$-independent and denote $\Omega = z I_H \langle T^z \rangle$.
In a low-density regime, we can focus only on the $n^a_i = 1$ subspace [see
Fig. \ref{fig:fig1}(d)].
There are two states with total spin projection $T^z_t = \pm 1$: $\vert t
\rangle_i = a^\dag_{i \ua} \vert \ua \rangle_i$ and $\vert \da \da \rangle_i =
a^\dag_{i \da} \vert \da \rangle_i$ and energies $E_{1, 2} = -\frac{\eta}{4} V
\mp \frac{1}{2} \Omega$; and two $T^z_t = 0$ states: $\vert u, s \rangle_i =
r_\pm a^\dag_{i \ua} \vert \da \rangle_i \mp r_\mp a^\dag_{i \da} \vert \ua
\rangle_i$ with energies $E_{u, s} = \frac{\eta}{4} V \pm \frac{1}{2}
\sqrt{V^2 + \Omega^2}$, where $\vert n \rangle$ is a core-spin state, $r_\pm =
\frac{1}{\sqrt{2}} (\cos \vartheta \pm \sin \vartheta)$, ${\rm tg} \, 2
\vartheta = \Omega / V$ (here $\vert \sigma \rangle_i$ is a shorthand notation
for $\vert \sigma \rangle_i \otimes \vert 0 \rangle$, $\vert 0 \rangle$ is the
$a$-fermion vacuum).
When $\Omega = \Omega_0 = (1 - \eta^2) V / 2 \eta$, $\vert s \rangle_i$ and
$\vert t \rangle_i$ become degenerate, and at strong-coupling define the local
s-t subspace.
For $\Omega \sim \Omega_0$ other states, separated by a large gap $\sim
\Omega_0$, can be ignored.
We represent this Hilbert space with {\it constrained (no double occupancy)
fermions} \cite{batista-2004-1}
\begin{equation}
 \dd_{i s} \vac \leftrightarrow \vert s \rangle_i, \,\,
 \dd_{i t} \vac \leftrightarrow \vert t \rangle_i.
 \label{eq:d-fermions}
\end{equation}
Here $\vac = \prod_i \vert \ua \rangle_i$ is a vacuum state with $n^a_i = 0$.

Near the resonance $\Omega = \Omega_0$, the system is described by an effective
Dirac-like Hamiltonian
\begin{equation}
 H_d \! = \! -J \! \sideset{}{_{\langle i j \rangle}}\sum
 [\sigma^x_{\alpha \beta} \dd_{i \alpha} d_{j \beta} \! + \! {\rm h.c.}] +
 \Delta \! \sideset{}{_i}\sum (n^d_{i s} - n^d_{i t}),
 \label{eq:dirac-model}
\end{equation}
obtained by projecting the Hamiltonian \eqref{eq:fm-xxz-klm} onto s-t subspace
\eqref{eq:d-fermions}, i.e. by computing matrix elements of Eq.
\eqref{eq:fm-xxz-klm} between states $\dd_{i \alpha} \vac$ \footnote{
 See the Supplementary Material, which includes Refs. \cite{cazalilla-2014-1,
 celi-2014-1,mancini-2015-1,stuhl-2015-1,fallani-2005-1,katori-2003-1,
 campbell-2008-1,safronova-2015-1,gradshteyn-2014-1,schollwock-2011-1,
 vidal-2004-1,white-2004-1,daley-2004-1}, for technical details of this
 calculation.
}.
In Eq. \eqref{eq:dirac-model}, $J = J_0 r_+$, $\alpha, \beta = s$ or $t$, and
$n^d_{i \alpha} = \dd_{i \alpha} d_{i \alpha}$ (with $n^d_i = n^d_{i s} +
n^d_{i t}$).
The first term contains hopping processes [see inset in Fig. \ref{fig:fig1}(d)]
that mix local entangled $\vert s \rangle$ and classical $\vert t \rangle$
states \eqref{eq:d-fermions} [because of $\sigma^x_{\alpha \beta}$].
This emergent spin-orbit coupling is rooted in an interplay between strong
exchange interactions and quantum fluctuations, and manifests in a {\it
transverse} [orthogonal to N\'eel vector $\langle S^z_i \rangle$] spin
polarization of $d$-particles: ${\bm T}^\perp_i = \frac{r_+}{r_-} {\bm
s}^\perp_{a i} = \frac{1}{2} r_+ \bsig^\perp_{\alpha \beta} \dd_{i \alpha}
d_{i \beta}$.
The second term contains an effective detuning $\Delta = \Omega - \Omega_0$
from the s-t resonance and describes a competition between Hund interaction and
AF correlations, both favoring an insulator (at large $\vert \Delta \vert \gg
J$) with localized fermions.
Remarkably, for $\vert \Delta \vert \sim J$ the state of the system is driven
by a subdominant kinetic-energy scale $J$, which stabilizes a {\it correlated
metallic phase} of $d$-fermions with transverse spin excitations.
Within the low-energy model \eqref{eq:dirac-model}, the sign of $\Delta$ is
irrelevant because under a canonical transformation $d_{i \alpha} \to
\sigma^x_{\alpha \beta} d_{i \beta}$, $H_d (-\Delta) \to H_d (\Delta)$.
Hence, below we assume that $\Delta \geqslant 0$.

The s-t resonance occurs because $\eta < 1$.
In the isotropic ($\eta = 1$) strong-coupling case Eq. \eqref{eq:fm-xxz-klm}
describes an insulator with localized triplets (similar to a DE model
\cite{anderson-1955-1}).
However, this state is unstable for $\eta < 1$ and $I_H \! \sim \! V$.
The existence of a s-t resonance does not contradict the ``poor man'' scaling
\cite{anderson-1970-1,georges-2013-1} where the $XY$ exchange $V$ flows to zero
at low energies.
Indeed, the latter is applicable only at weak coupling $V \ll J_0$, while our
theory operates in the opposite, {\it strong coupling regime} $V \gg J_0$.

\paragraph*{Singlet-triplet resonance with AEAs.--}

The correlated metal phase may be challenging to observe in a solid-state
system due to a multitude of parameters in Eq. \eqref{eq:fm-xxz-klm} that need
to be tuned near the s-t resonance.
Here we propose to realize this phase using two-level AEAs in an optical
lattice of Fig. \ref{fig:fig1}(c).
The spin-$\frac{1}{2}$ degrees of freedom can be implemented using either (i)
nuclear spins of atoms in the $\vert g \rangle$ electronic state, or (ii)
$\vert e, g \rangle$ clock states of nuclear-spin polarized atoms.
The Hamiltonian of the system is:
\begin{align}
 H = & -J_0 \! \sum_{\langle i j \rangle} (\cd_{i 2 n} c_{j 2 n} +
 {\rm h.c.}) - \sum_i \bigl[ V \cd_{i 1 n} c_{i 1 m} \cd_{i 2 m} c_{i 2 n} +
 \nonumber \\
 + & \Omega (-1)^i (\cd_{i a \ua} c_{i a \ua} - \cd_{i a \da} c_{i a \da})
 \bigr] + U \sum_i n^c_{i 2 \ua} n^c_{i 2 \da},
 \label{eq:klm}
\end{align}
where $\cd_{i a n}$ creates a fermion at site $i$ in Bloch band $a = 1, 2$ with
spin $n$.
Band 1 is localized and contains one atom per site, while the itinerant band 2
has an arbitrary filling and a NN hopping $J_0$ \footnote{
 In a 2D optical lattice, the hopping in an excited band generally depends on
 direction due to different motional states.
 This complication may be avoided, if uses the third excited band for mobile
 atoms.
}.
The second term in Eq. \eqref{eq:klm} is an interband exchange interaction.
It is FM ($V > 0$) because atoms can experience $s$-wave collisions only in a
spin-singlet state.
The third term contains the {\it staggered} [indicated by $(-1)^i =
\e^{\ii {\bm Q} \cdot {\bm x}_i}$] Zeeman-like Raman [in case (i)] or direct
Rabi [for case (ii)] coupling, with $\Omega > 0$ which simulates the AF
environment in Fig. \ref{fig:fig1}(b) \footnote{
 See the Supplementary Material for further details of the cold-atom
 implementation \eqref{eq:klm}.
}.
Finally, there is a local repulsion $U$ due to intraband $s$-wave collisions
($n^c_{i a n} = \cd_{i a n} c_{i a n}$).

Since atoms in band 1 are localized they only contribute spin degrees of
freedom, $\bS_i = \frac{1}{2} \bsig_{n m} \cd_{i 1 n} c_{i 1 m}$.
Up to a density-density interaction, magnetic terms in Eq. \eqref{eq:klm} can
be rewritten as $-2 \sum_i \bigl[ V \bS_i \cdot {\bm s}_{c i} + \Omega (-1)^i
(S^z_i + s^z_{c i}) \bigr]$, where we omitted the band index, $c_{i n} \equiv
c_{i 2 n}$.
Applying the transformation \eqref{eq:st-trans} to Eq. \eqref{eq:klm}, we get
rid of $(-1)^i$, and replace $c_{i n} \to a_{i n}$ (${\bm s}_{c i} \to
{\bm s}_{a i}$) and $\bS_i \to {\bm T}_i$.

On an isolated site $i$ there are $8$ eigenvalues $E_{n^a} (T_t, T^z_t)$
labeled by the total spin $T_t$, its projection $T^z_t$ and fermion number
$n^a$: $E_0 \bigl( \frac{1}{2}, \pm \frac{1}{2} \bigr) = E_2 \bigl(
\frac{1}{2}, \pm \frac{1}{2} \bigr) - U = \mp \Omega$, $E_1 (0, 0) = - 3 E_1
(1, 0) = \frac{3}{2} V$ and $E_1 (1, \pm 1) = -\frac{V}{2} \mp 2 \Omega$.
Energy levels with $n^a = 1$ are shown in Fig. \ref{fig:fig1}(e).
For small $J_0$ a mobile atom can propagate only when two or more states are at
resonance, i.e. for $\Omega = 0$ or $V$.
The first case is a usual FKLM \cite{izyumov-2001-1} without AF correlations.

We will concentrate on the second resonance at $\Omega = V$, reached in an {\it
excited} s-t manifold spanned by the local singlet $\vert s \rangle_i =
\frac{1}{\sqrt{2}}[a^\dag_{i \ua} \vert \da \rangle_i - a^\dag_{i \da} \vert
\ua \rangle_i]$ and triplet $\vert t \rangle_i = a^\dag_{i \da} \vert \da
\rangle_i$ [red circle in Fig. \ref{fig:fig1}(e)], and identify these states
with the corresponding states \eqref{eq:d-fermions}: now $\dd_{i s}$ creates a
pure spin-singlet, while $\dd_{i t}$ creates a triplet.
In an excited manifold, the vacuum state $\vac = \prod_i \vert \da \rangle_i$
has local spins {\it antiparallel} to $\Omega$.
Near the s-t resonance, other singly-occupied states are separated by a gap
$\sim V$ and can be ignored, together with the doubly-occupied manifold.
The system is described by the effective model \eqref{eq:dirac-model} with $J =
\frac{1}{\sqrt{2}} J_0$, $\Delta = V - \Omega$ \footnote{
 See the Supplementary Material for a detailed explanation when one can ignore
 the doubly-occupied states, and for a calculation of second-order corrections
 $\sim J_0^2 / \Omega$.
}.
Thus, in a strong-coupling regime $J_0 \ll \Omega$, $V$ and for $\Omega \sim
V$, the AEA setup \eqref{eq:klm} can be used to simulate the s-t resonance
dynamics of Eq. \eqref{eq:fm-xxz-klm}.
For example, the transverse magnetization of a $d$-particle is now
${\bm T}^\perp_i = -{\bm s}^\perp_{a i} = -\bsig^\perp_{\alpha \beta}
\dd_{i \alpha} d_{i \beta} / \sqrt{8}$.
We focus on the excited s-t manifold because a cold-atom system is usually
well-isolated from its environment and can not escape the s-t subspace due to
decoherence.

\begin{figure}[t]
 \begin{center}
  \includegraphics[width = \columnwidth]{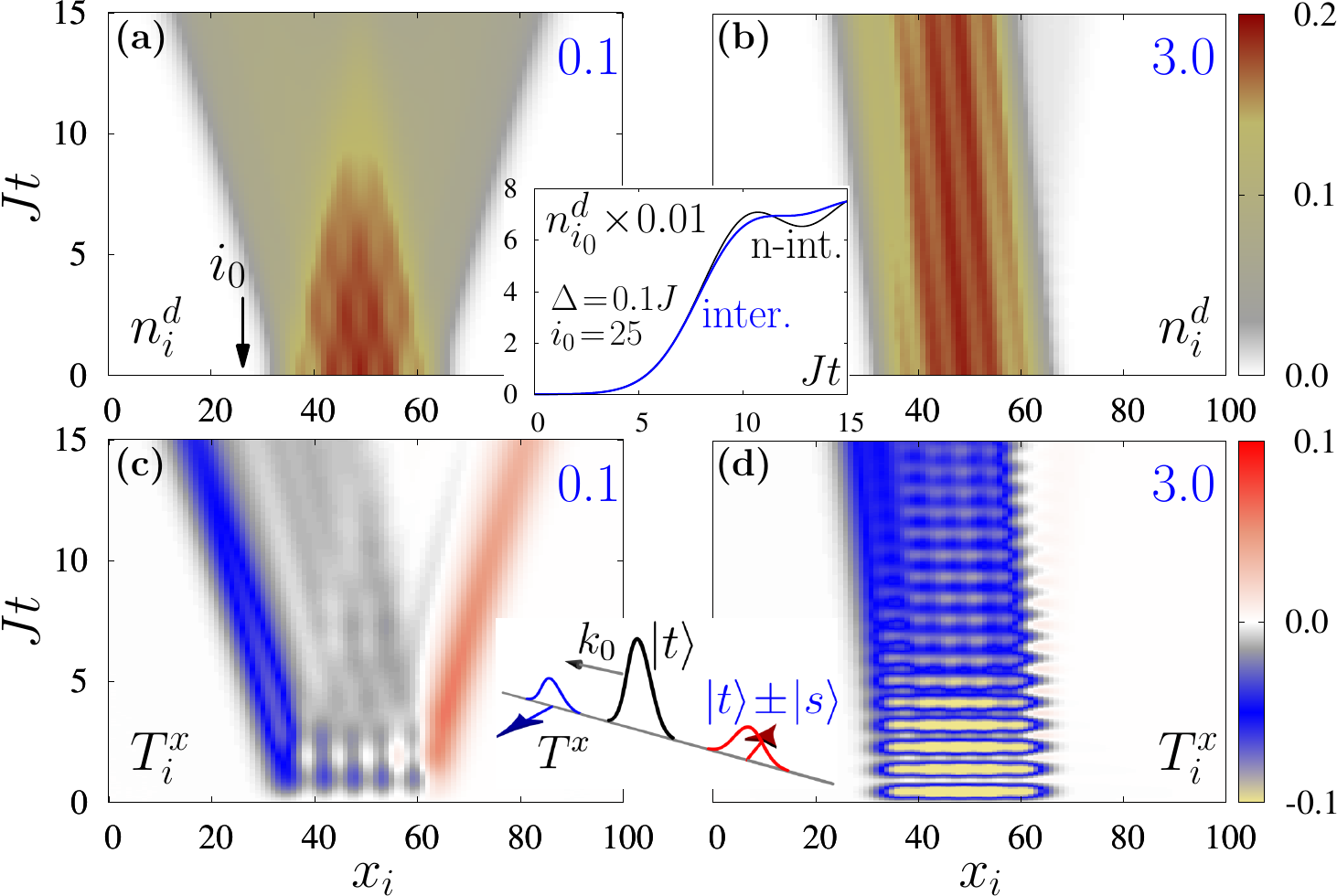}
 \end{center}
 \caption{
  Propagation of wavepackets with 5 particles.
  {\bf (a)} and {\bf (b)} show total density $\langle n^d_i (t) \rangle$;
  {\bf (c)} and {\bf (d)} contain the transverse local-spin polarization
  $\langle T^x_i \rangle$.
  Blue numbers in top right corners indicate the detuning $\Delta / J$.
  Inset: comparison between density evolution at a fixed $x_i = i_0$ [black
  arrow in panel (a)] in the full model \eqref{eq:dirac-model} (blue line) and
  in the case of non-interacting (n-int) $d$-fermions (black line).
  The parameters are $N = 101$, $A = 10^{-3} J$, and $k_0 = -20 \pi / N$.
 }
 \label{fig:fig2}
\end{figure}

\paragraph*{Wavepacket dynamics in 1D.--}

The spin-motion coupling and transverse spin of $d$-fermions can be probed via
propagation of many-body wavepackets.
We focus on a 1D case and study dynamics of the model \eqref{eq:dirac-model}
within a time-dependent density matrix renormalization group (t-DMRG) method
\cite{schollwock-2011-1,vidal-2004-1,white-2004-1,daley-2004-1}.
The initial wavefunction is assumed to contain only triplets and is a ground
state (GS) of the Hamiltonian $H_{\rm 1 D} (t < 0) = -J \sum_i \bigl[ \dd_{i t}
d_{i + 1, t} + {\rm h.c.} \bigr] + A \sum_i \bigl[ x_i - \frac{N}{2} \bigr]^2
n^d_{i t}$ that describes fermions $d_{i t}$ in a harmonic trap with $A > 0$.
At $t = 0$, the trap is removed, so $H_{\rm 1 D} (t \geqslant 0) = H_d$, and
the packet is accelerated to a momentum $k_0$ by applying an operator
$\e^{\ii k_0 \sum_i x_i n^d_{i t}}$.

Fig. \ref{fig:fig2} shows evolution of five-fermion wavepackets for $\Delta =
0.1 J$ and $\Delta = 3 J$.
Near the s-t resonance, the initial distribution splits into two fast
counter-propagating parts with {\it opposite} transverse local magnetization
$\langle T^x_i \rangle$ [Fig. \ref{fig:fig2}(a) and (c)], while for large
$\Delta$ this splitting is negligible and the state remains practically
localized [Fig. \ref{fig:fig2}(b) and (d)].
To understand this behavior, we consider dynamics of a single $d$-fermion, when
the Hamiltonian \eqref{eq:dirac-model} can be diagonalized in terms of
quasiparticles with dispersion $\varepsilon_{k \lambda} = \lambda \rho_k =
\lambda \sqrt{(2 J \cos k)^2 + \Delta^2}$ ($\lambda = \pm$) \footnote{
 See the Supplementary Material for dynamics of single-particle wavepackets.
}.
Because these bands have opposite group velocities $v_k^\lambda = \lambda
\partial_k \rho_k$, after a time $t > 1 / J$ the density has an approximate
form $\langle n^d_{i \alpha} (t) \rangle \approx R_\alpha (\xi_-) + L_\alpha
(\xi_+)$ where $\xi_\pm = x_i \pm v_{k_0}^- t$, $v_{k_0}^- = -2 J \vert \sin
k_0 \vert / \rho_{k_0}$, and $L_\alpha$, $R_\alpha$ with $\frac{R_s}{L_s}
\approx 1$ and $\frac{R_t}{L_t} \approx \bigl( \frac{\rho_{k_0} -
\Delta}{\rho_{k_0} + \Delta} \bigr)^2$ describe right and left movers.
For large $\Delta$, $R_t \ll L_t$ [see Fig. \ref{fig:fig2}(b)].
Similarly, $\langle T^x_i (J t > 1) \rangle \approx R(\xi_-) + L(\xi_+)$, with
$\frac{R}{L} \approx -\frac{\rho_{k_0} - \Delta}{\rho_{k_0} + \Delta} < 0$
[Fig. \ref{fig:fig2}(c)].
This single-particle picture is valid near wavepacket edges with low fermion
density [see inset of Fig. \ref{fig:fig2}], and breaks down at the
strongly-correlated core.

\begin{figure}[t]
 \begin{center}
  \includegraphics[width = \columnwidth]{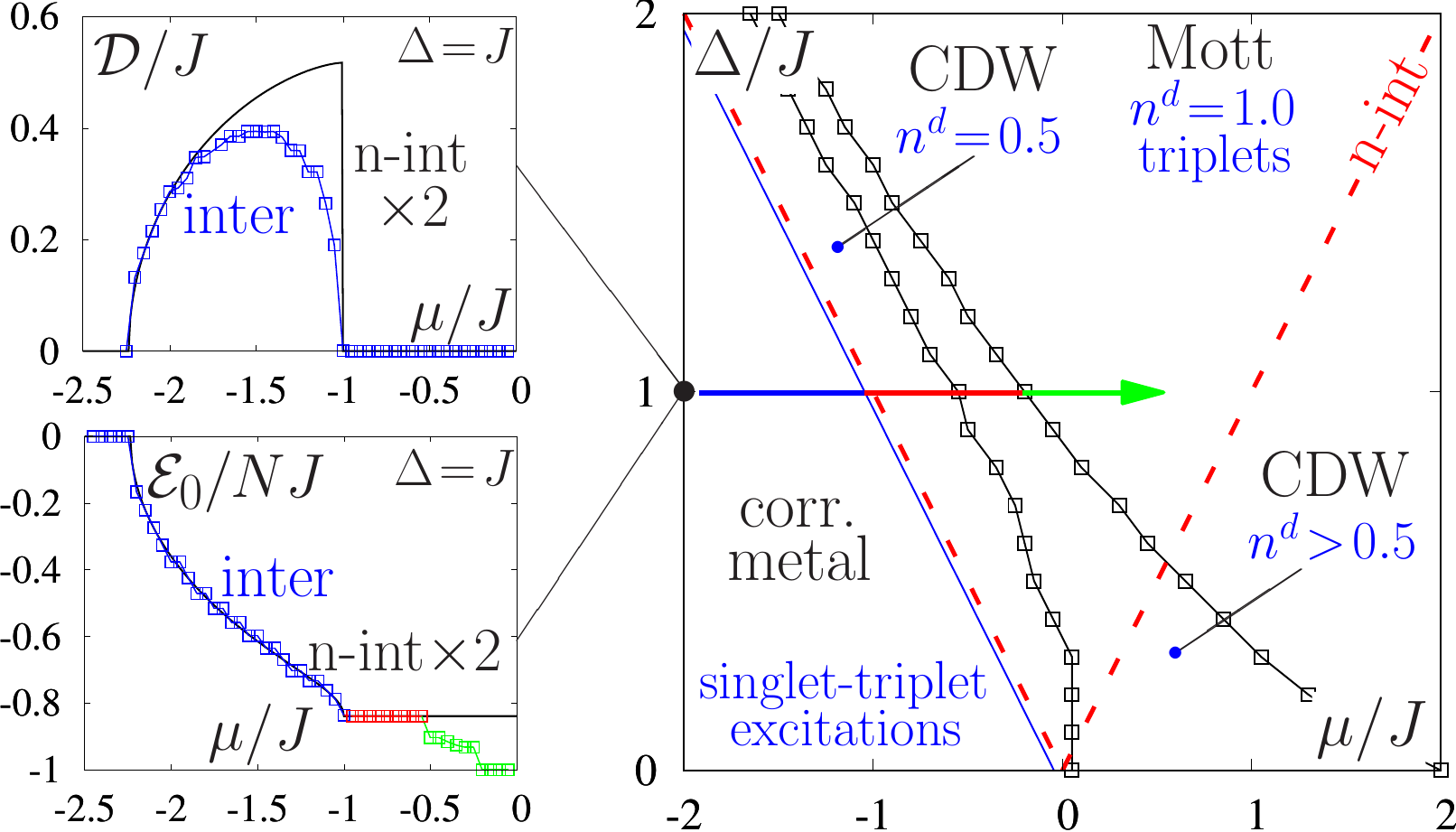}
 \end{center}
 \caption{
  Phase diagram of the model \eqref{eq:dirac-model} on a $N = 40$ site chain
  with periodic boundary conditions.
  Black and blue lines mark 1st order transitions.
  For the correlated metal phase, the Drude weight ${\cal D} > 0$, while in all
  other states ${\cal D} = 0$.
  Dashed red line separates metallic (below) and band insulator (above) states
  in a model with non-interacting (n-int) $d$-fermions.
  At $\Delta = 0$, ${\cal D} = \frac{1}{\pi} \sqrt{(2 J)^2 - \mu^2}$ for $\vert
  \mu \vert < 2 J$ and $0$ otherwise.
  On the left: Drude weight and ground-state energy ${\cal E}_0$ for $\Delta =
  J$ plotted along the arrow in the main panel.
  Notations are as in Fig. \ref{fig:fig2}.
  Notice jumps in $\frac{\partial {\cal E}_0}{\partial \mu}$ at phase
  transitions.
  For $\mu \approx -0.5$ (CDW state), ${\cal E}_0$ decreases with
  increasing $n^d_t$.
  For n-int fermions, one has to multiply ${\cal D}$ and ${\cal E}_0$ (black
  curves) by 2 due to spin degeneracy, absent for constrained $d$-fermions.
 }
 \label{fig:fig3}
\end{figure}

\paragraph*{The correlated metallic state.--}

To capture interaction effects that lead to correlated metal phase and drive
metal-insulator transitions, in Fig. \ref{fig:fig3} we compute phase diagram of
Eq. \eqref{eq:dirac-model} in 1D, as a function of the detuning $\Delta$ and
chemical potential $\mu$ (described by a term $\delta H_d = -\mu \sum_i
n^d_i$).
We characterize various GSs with a Drude weight (DW) ${\cal D}$ related to the
longitudinal conductivity as ${\rm Re} \, \sigma_{x x} (\omega \to 0) =
{\cal D} \delta (\omega)$: ${\cal D} > 0$ for a metal and vanishes in an
insulator.
For a system with periodic boundary conditions, ${\cal D} = \frac{1}{N}
\frac{d^2 {\cal E}_0}{d \phi^2} \bigl \vert_{\phi = 0}$ where ${\cal E}_0$ is
the GS energy and $\phi N$ is the flux piercing the ring
\cite{kohn-1964-1,scalapino-1993-1}.
In Eq. \eqref{eq:dirac-model} we replace $\dd_{i \alpha} d_{i + 1, \beta} \to
\dd_{i \alpha} \e^{\ii \phi} d_{i+1, \beta}$ and treat this model using an
unbiased DMRG technique \footnote{
 See the Supplementary Material for details of DMRG simulations.
}.

The physics of a non-interacting (n-int) model \eqref{eq:dirac-model} is
determined by filling of single-particle bands $\varepsilon_{k \lambda}$: When
$\mu$ is inside one of them the system is a metal (regions below dashed red
line in Fig. \ref{fig:fig3}), otherwise it is a band insulator
\footnote{
 See the Supplementary Material for derivation of the Drude weight ${\cal D}$
 for non-interacting $d$-fermions.
}.
The correlated nature of $d$-fermions qualitatively changes this picture by
dramatically suppressing the metallic phase and transforming the band insulator
to either {\it charge density-wave} (CDW) with $n^d < 1$, or a {\it triplet
Mott state} with $n^d = n^d_t = 1$.
Metallic, CDW and Mott phases are separated by a 1st order transition.
Surprisingly, a CDW state with $n^d = 0.5$ emerges exactly at $\mu = - \Delta$,
right at the metal-insulator transition for non-interacting fermions.
While the latter is driven by a simple band filling, the CDW arises due to
quantum effects: For $\mu = - \Delta$ the on-site energy of a triplet vanishes
which results in a macroscopic degeneracy (associated with different fillings
of triplets) of the classical GS.
Quantum fluctuations due to s-t virtual hoppings then select a GS with a
two-site unit cell.
For $n^d > 0.5$ this state evolves into CDWs with larger unit cells.

At low density, one can extract the DW from a group velocity of a wavepacket
with small momentum $k_0$, $v_{k_0}^- \approx \frac{k_0}{m^*}$ [$m^* \!\! = \!
\! \frac{1}{(2 J)^2} \sqrt{(2 J)^2 \! + \! \Delta^2}$], as ${\cal D} (n^d \!
\ll \! 1) \! \approx \! \frac{n^d}{m^*} = \frac{n^d v_{k_0}^-}{k_0}$.

\paragraph*{Discussion.--}

The state of a many-body system with competing strong interactions often has
unexpected physical properties and is driven by a subdominant energy scale.
We illustrated this mechanism in a FKLM where an interplay between strong
on-site FM exchange and AF correlations, each favoring an insulating behavior,
allows the small kinetic energy to stabilize a correlated {\it
metallic} phase, whose elementary excitations involve resonating singlet and
triplet states of bare local spins and mobile fermions.
This s-t mixing leads to a distortion of the AF order and local magnetization
{\it perpendicular} to the N\'eel vector.
We also showed how one can probe this physics in a quantum simulator with AEAs
in optical lattices under a laser-induced magnetic field.

Our results, obtained using a low-energy model \eqref{eq:dirac-model}, remain
valid within the full Hamiltonian
\eqref{eq:klm} with $\Omega > J$ \footnote{
 See the Supplementary Material for a comparison of the wavepacket dynamics
 in these two cases.
},
and should be applicable beyond 1D, because the phases in Fig. \ref{fig:fig3}
are not associated with spontaneous breaking of any continuous symmetry.

The observation of wavepacket dynamics in Fig. \ref{fig:fig2} and transverse
spin excitations does not require temperatures $\sim J$ and relies on an
uncorrelated initial triplet state.
These features can be probed in quench experiments with AEAs in moving optical
lattices \cite{mun-2007-1}.
The Drude weight ${\cal D}$ can be measured as a response to a weak optical
lattice tilt \cite{raizen-1997-1}.
Thus the phase diagram in Fig. \ref{fig:fig3} can be verified, at least for
low-densities and deep lattices when the band relaxation due to
collisions is energetically suppressed.

\paragraph*{Acknowledgments.--}

We thank Ivar Martin for illuminating discussions.
This work was supported by NSF (PHY-1211914, PHY-1521080 and PFC-1125844),
AFOSR FA9550-13-1-0086, AFOSR-MURI Advanced Quantum Materials, NIST and ARO
W911NF-12-1-0228 individual investigator awards.

\bibliographystyle{apsrev4-1}
\bibliography{lit}
\end{document}


\title{Supplementary material for: \\
 ``Spin-orbit coupled correlated metal phase in Kondo lattices: an
 implementation with alkaline-earth atoms''}
\author{}
\affiliation{}
\begin{abstract}
\end{abstract}
\pacs{}
\maketitle

\section{Derivation of the model in Eq. (4)}

The effective low-energy model (4) was obtained by projecting the full
Hamiltonian (1) of a solid-state FKLM [or its AMO analog in Eq. (5)] onto the
s-t subspace (3).
In this section, we describe details of this derivation.

The s-t manifold (3) is spanned by the states $\vert s \rangle_i = r_-
a^\dag_{i \ua} \vert \da \rangle_i + r_+ a^\dag_{i \da} \vert \ua \rangle_i$
and $\vert t \rangle_i = a^\dag_{i \ua} \vert \ua \rangle_i$ that diagonalize
{\it local magnetic interactions} in the Hamiltonian (1) [see discussion after
Eq. (2)].
These local terms contribute a detuning from the s-t resonance
\begin{displaymath}
 H_{\rm loc} \to \Delta \sum_i \bigl[ \vert s \rangle_i \langle s \vert_i -
 \vert t \rangle_i \langle t \vert_i\bigr].
\end{displaymath}
Hence, we only need to compute matrix elements in the s-t subspace of the
kinetic energy from Eq. (1).
In the staggered frame (2), it has the form $H_0 = -J_0
\sigma^x_{n m} \sum_{\langle i j \rangle} (\ad_{i n} a_{j m} + {\rm h.c.})$.
Because a given lattice site can not be simultaneously occupied by a singlet
and a triplet (no double occupancy constraint), non-zero matrix elements will
be between states of the type $\vert \alpha \rangle_i \vert \ua \rangle_j$ and
$\vert \ua \rangle_i \vert \beta \rangle_j$ with $\alpha$, $\beta = s$ or $t$,
and nearest-neighbor (NN) sites $i$ and $j$.
We have:
\begin{align}
 H_0 \vert s \rangle_i \vert \ua \rangle_j = & -J_0 \bigl[ {\color{blue}
 \cancel{\color{black} r_- \vert \da \rangle_i \ad_{j \da} \vert \ua
 \rangle_j}} + r_+ \vert \ua \rangle_i \vert
 t \rangle_j \bigr] \to \nonumber \\
 & \to -J_0 r_+ \vert \ua \rangle_i \vert t \rangle_j, \nonumber \\
 H_0 \vert t \rangle_i \vert \ua \rangle_j = & -J_0 \vert \ua \rangle_i
 \ad_{j \da} \vert \ua \rangle_j = \nonumber \\
 & = -J_0 \vert \ua \rangle_i \bigl[ r_+ \vert s \rangle_j - {\color{blue}
 \cancel{\color{black} r_- \vert u \rangle_j}} \bigr] \to J_0 r_+ \vert \ua
 \rangle_i \vert s \rangle_j. \nonumber
\end{align}
The crossed terms are off-resonant, i.e. do not belong to the local s-t
manifold.
Combining $H_0$ and $H_{\rm loc}$, we obtain $H_d$, Eq. (4), with $J = J_0
r_+$.

For the AEA Hamiltonian (5), the s-t subspace is defined by the local states
$\vert s \rangle_i = \frac{1}{\sqrt{2}} [\ad_{i \ua} \vert \da \rangle_i -
\ad_{i \da} \vert \ua \rangle_i]$ and $\vert t \rangle_i = \ad_{i \da} \vert
\da \rangle_i$.
Performing same steps as above, we arrive at the model (4) with $J = J_0 /
\sqrt{2}$.

\section{Alkaline-earth atoms in a laser-induced magnetic field}

This section is dedicated to general remarks regarding the setup with AEAs in
an optical lattice shown in Fig. 1(d) and described by Eq. (5).
Among other issues, we will clarify the FM nature of atomic exchange
interactions, the origin of the artificial Zeeman field $\Omega$, and justify
our focus on the singlet-triplet resonance that played a central role in the
main text.

\subsection{Exchange interactions and artificial Zeeman field}

Alkaline-earth fermionic atoms, such as ${}^{171} {\rm Yb}$,
${}^{173} {\rm Yb}$ and ${}^{87} {\rm Sr}$, have two valence electrons in a
state with total angular momentum $J_t = 0$, and a finite nuclear spin $I$ ($I
= 1 / 2$, $5 / 2$ and $9 / 2$ for ${}^{171} {\rm Yb}$, ${}^{173} {\rm Yb}$ and
${}^{87} {\rm Sr}$, respectively) \cite{cazalilla-2014-1}.
Additionally, there are two electronic ''clock'' states: lowest-energy orbital
singlet ${}^1 S_0$ ($\vert g \rangle$) and an excited triplet ${}^3 P_0$
($\vert e \rangle$).
Because $\vert e \rangle$ and $\vert g \rangle$ configurations have $J_t = 0$,
they are almost perfectly decoupled from the nuclear degrees of freedom during
$s$-wave collisions, which allows us to use either electronic or nuclear-spin
states to encode pseudo-spin flavors $n = \ua$, $\da$.
At the level of only $s$-wave two-atom interactions, both choices yield the
FM form of the exchange coupling in Eq. (5).

Indeed, suppose that all atoms are nuclear spin-polarized and pseudo-spins
$n = \ua, \da$ are identified with clock states $\vert e, g \rangle$.
Since the two atoms reside in two different Bloch bands [lowest $\vert 1
\rangle$ and excited $\vert 2 \rangle$, see Fig. 1(d)], we must antisymmetrize
their total, i.e. spatial and electronic, wavefunction.
If we focus only on $s$-wave two-body collisions, the spatial wavefunction
must be symmetric and electronic part -- antisymmetric, so only $e g$-singlets
can scatter.
Because the corresponding scattering length $a_{e g}^-$ is positive, the $e g$
two-body singlet has higher energy than the triplet, leading to a FM exchange
$V < 0$ (which favors $e g$-triplets with zero energy).

\begin{figure}[t]
 \begin{center}
  \includegraphics[width = \columnwidth]{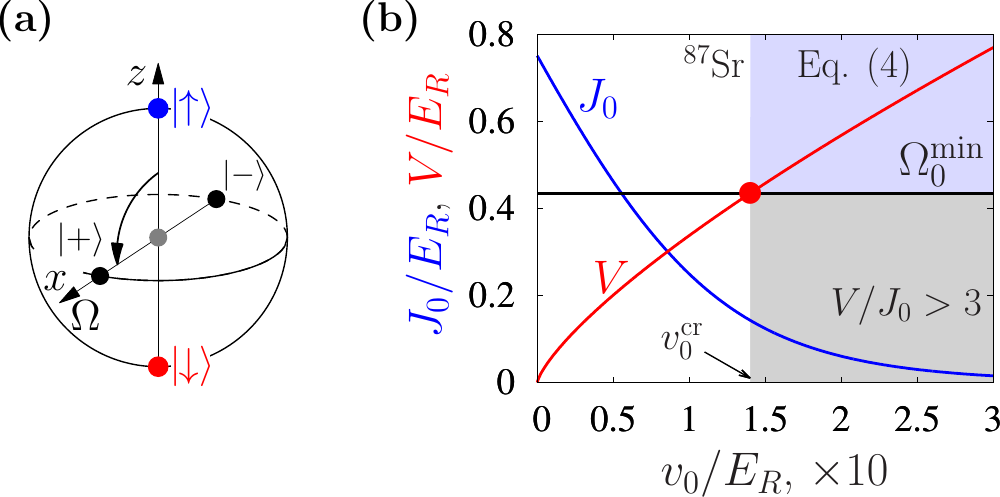}
 \end{center}
 \caption{
  {\bf (a)} Rotation that aligns the $z$-axis in pseudo-spin space with the
  direction of the artificial magnetic field $\Omega$.
  {\bf (b)} Excited-band hopping $J_0$ and interband exchange interaction $V$
  in the Sr optical lattice clock as functions of the lattice depth $v_0$.
  The photon recoil energy is $E_R \approx 3.2 \, {\rm kHz}$.
  The red dot marks the lowest value of the Rabi coupling at a s-t resonance
  $\Omega^{\rm min}_0 = V (v_0^{\rm cr})$ when $V / J_0 = 3$.
  Gray shading shows the range of lattice depths with $V / J_0 > 3$.
  When in addition $\Omega > \Omega_0^{\rm min}$ (light blue region) the
  effective Hamiltonian (4) is a good approximation to the full model (5).
 }
 \label{fig:sfig1}
\end{figure}

These arguments remain essentially unchanged in the other case when
pseudo-spins $\ua$ and $\da$ are associated with any two nuclear spin states.
In this case, we assume that all atoms are prepared in the lowest clock
configuration $\vert g \rangle$.
As before, the spatial part of the wavefunction must be symmetric, so the only
scattering channel is nuclear-spin singlet with a positive scattering length
$a_{g g}$.
Thus, the exchange $V$ is again FM, and our analysis in the main text covers
both above cases.

In the pseudo-spin language, the FM exchange interaction has full $SU (2)$
symmetry, see Eq. (5).
This fact is a consequence of the decoupling between electronic and
nuclear-spin degrees of freedom \cite{cazalilla-2014-1}, but can also be
understood by recalling that only pseudo-spin {\it singlets} participate in
$s$-wave collisions.
The $SU (2)$ symmetry of interactions allows us to choose the spin quantization
axis arbitrarily, and simplify the artificial magnetic field term.

The latter is laser-induced by coupling internal atomic levels via direct
optical (if pseudo-spins correspond to electronic $e$-$g$ states) or Raman
two-photon (when pseudo-spins are encoded in nuclear spins) transitions
\cite{celi-2014-1,mancini-2015-1,stuhl-2015-1}.
This coupling is equivalent to a magnetic field along the $x$-axis: $H_R =
\Omega \, \sigma^x_{a b} \vert a \rangle \langle b \vert$ with $a$ and $b =
\ua$, $\da$.
We can use the $SU (2)$ symmetry and rotate the basis to align the $z$-axis
with this field, see Fig. \ref{fig:sfig1}: $H_R = \Omega \, \sigma^z_{a b}
\vert a \rangle \langle b \vert$.

\subsection{Realizing a staggered artificial magnetic field}

The quantum simulator AEA setup in Eq. (5) relies on a staggered nature of the
laser-induced artificial magnetic field that has opposite signs on NN lattice
sites.
Naturally, an implementation of this $\pi$-modulated field depends on the
physical degrees of freedom used to encode pseudo-spin flavors.

First, let us assume that pseudo-spins $\ua, \da$ correspond to nuclear-spin
states (with all atoms in the $\vert g \rangle$ electronic configuration) and
are coupled via two-photon Raman transitions \cite{mancini-2015-1,stuhl-2015-1}
using retroreflected laser beams with wavevectors $\bk_1$ and $-\bk_2$.
In this setup, spatial dependence of the laser-induced field is given by a
standing wave $\Omega \e^{\ii (\bk_1 - \bk_2) \cdot {\bm x}_i} + {\rm c.c.}$
(${\bm x}_i$ labels optical lattice sites, and ``c.c.'' stands for
``complex-conjugate'').
In 1D, ${\bm x_i} = {\bm e}_0 a_0 i $ with $a_0$ being the lattice spacing and
${\bm e}_0$ -- a unit vector along the lattice that forms an angle $\alpha$
with $\bk_1 - \bk_2$.
One way to achieve a staggered artificial field, is to tune this angle so that
$\cos \alpha = \pi / a_0 \vert \bk_1 - \bk_2 \vert$.
Alternatively, one can simply adjust the wavelength of the optical lattice
potential $\lambda_0 = 2 a_0$  to match the wavelength of a laser that imprints
the staggered phase.
This can be done by tuning the relative angle $\vartheta$ between lattice laser
beams \cite{fallani-2005-1} as $a_0 = \lambda_0 / 2 \sin \frac{\vartheta}{2}$.

The situation is slightly different when pseudo-spin components are encoded
with clock $\vert e \rangle$ and $\vert g \rangle$ states, because the optical
lattice wavelength must be magic, i.e. chosen in such a way that electronic
polarizabilities of both electronic configurations coincide
\cite{katori-2003-1}.
This magic wavelength, $\lambda_0^m$, needs to be small enough compared to the
wavelength $\lambda_c$ of the $e$-$g$ transition, so that the equality
$(2 \pi / \lambda_c) a_0^m = \pi \lambda_0^m / \lambda_c \sin
\frac{\vartheta}{2} = \pi$ or $\sin \frac{\vartheta}{2} = \lambda_0^m /
\lambda_c$ can be satisfied for some value of $\vartheta$.
For example, for ${}^{87} {\rm Sr}$, $\lambda_c = 698 \, {\rm nm}$
\cite{campbell-2008-1} and there are five magic wavelengths
\cite{safronova-2015-1}: one at $\lambda_0^m = 813 \, {\rm nm}$ and four with
$\lambda_0^m < \lambda_c$.
Using any of the latter for optical lattice lasers (plus a retroreflected
probing beam at the resonance with the $e$-$g$ transition) would realize the
staggered artificial magnetic field.

Another way of implementing the staggered artificial field is to load atoms in
a 3D anisotropic optical lattice where tunneling along one direction exceeds
hopping in the other directions.
In this geometry, by aligning the clock laser w.r.t. to the relevant tunneling
direction, one can ensure that it imprints a phase of $\pi$ on the atoms.

\subsection{Estimates of experimental parameters}

Having discussed general features of the proposed experimental setup, here we
estimate relevant energy scales required to realize the correlated metal state
in Figs. 1(c) and Fig. 3.
We assume that the experiment will use ${}^{87} {\rm Sr}$ atoms in a magic
optical lattice (wavelength $\lambda_m$) described by a potential
$V_{\rm las} ({\bm x}) = v_0 (\sin^2 k x + \sin^2 k y + \sin^2 k z)$ with $k =
2 \pi / \lambda_m$, and consider identical confinement along all directions.

In Fig. \ref{fig:sfig1}(b) we show lattice-depth ($v_0$) dependence of the
hopping amplitude $J_0$ between same spatial orbitals in an excited band, and
exchange interaction $V$ between the lowest and first excited bands.
The parameters of the system are: atomic mass $m_{\rm Sr} = 87 \, {\rm a.u.}$;
magic wavelength $\lambda_m = 813 \, {\rm nm}$ which translates into a recoil
energy $E_R = \frac{\hbar^2 k^2}{2 m_{\rm Sr}} \approx 3.2 \, {\rm kHz}$; and
scattering length $a = a_{g g} \sim a^-_{e g} \sim 200 \, a_0$ ($a_0$ is the
Bohr radius).
We compute the hopping $J_0$ by directly solving a 1D single-particle
Sch\"odinger equation in the periodic potential $V^{\rm 1D}_{\rm las} = v_0
\sin^2 k x$, as a quarter of the first excited band width.
To find the interband exchange, we employ a harmonic approximation
$V_{\rm las} \approx v_0 k^2 {\bm x}^2 \equiv \frac{1}{2} m_{\rm Sr} \omega^2
{\bm x}^2$ with $\omega = \frac{2}{\hbar} \sqrt{v_0 E_R}$.
Following the procedure explained in the next subsection and using the above
numerical values, one obtains $V (v_0) / E_R \approx 0.06 (v_0 / E_R)^{3 / 4}$.

We can use these results to determine experimental conditions under which the
effective low-energy model (4) is realized.
First, $V / J_0$ must be large.
As we show in the last section of this Supplementary Material, Eq. (4) provides
a satisfactory approximation to the full model (5) even for $V \gtrsim 2 J_0$
and small detuning $\Omega \sim V$.
Hence, we require $V / J_0 > 3$ [gray region in Fig. \ref{fig:sfig1}(b)] which
puts a lower bound on the lattice depth $v_0^{\rm cr} = 14 \, E_R$.

Next, the Rabi coupling $\Omega$ must be close to its value $\Omega_0 = V$ at
the s-t resonance [see Figs. 1(d) and 1(e)], in the sense that even for a
detuning $\Delta \sim J_0$, both $V$ and $\Omega$ remain larger than $J_0$.
The lower bound on $\Omega_0$ is given by $\Omega_0^{\rm min} = V
(v_0^{\rm cr}) \sim 1.5 \, {\rm kHz}$ [black line in \ref{fig:sfig1}(b)].
For deeper lattices with $v_0 > v_0^{cr}$, the low-energy model (4) should
become more accurate and offer access to a wider range of detunings $\Delta$,
provided a strong enough $\Omega > \Omega^{\rm min}_0$ can be realized [light
blue region in Fig. \ref{fig:sfig1}(b)].

\subsection{Role of the local intraband scattering (Hubbard repulsion)}

So far we assumed that the local repulsion $U$ in Eq. (5) can be ignored.
Now we will identify the parameter regime where this assumption is valid.

Let us first consider a single well and compute the energy of a state with
three particles: one in the lowest and two in the higher band.
There is only one relevant $s$-wave scattering energy, $u_s > 0$, that
corresponds to a pseudo-spin singlet state.
Depending on whether pseudo-spin degrees of freedom are implemented with
electronic $g$ and $e$ states of nuclear-spin polarized atoms, or with nuclear
spins of atoms in the $g$ clock state, $u_s = \frac{4 \pi}{m} a^-_{e g}$ or
$\frac{4 \pi}{m} a_{g g}$, respectively (we use the units with $\hbar = 1$).
The single-well effective Hamiltonian (omitting the site index $i$) can be
written as [cf. Eq. (5)]:
\begin{align}
 H_{\rm ex} = & \frac{U}{2} n^c_2 (n^c_2 - 1) + V \biggl[ \sum_\sigma
 n^c_{1 \sigma} n^c_{2, -\sigma} - \nonumber \\
 & - \bigl( \cd_{1 \ua} c_{1 \da} \cd_{2 \da} c_{2 \ua} + \cd_{1 \da} c_{1 \ua}
 \cd_{2 \ua} c_{2 \da} \bigr) \biggr], \nonumber
\end{align}
where $n^c_a = \sum_\sigma \cd_{a \sigma} c_{a \sigma}$, and indices $a = 1, 2$
denote lowest and excited bands (motional states).
Because the triplet state has zero energy, this Hamiltonian adds an energy
shift $V n^c_2$ to each on-site term in $H$ from Eq. (5).

To estimate a relative magnitude of $U$ and $V$, we assume a 1D harmonic
quantum well ${\cal V} (x) = \frac{1}{2} m \omega^2 x^2$ where only lowest $n =
0$ and excited $n = 1$ states are populated.
Then a simple calculation yields:
\begin{align}
 U = u^{\rm 1 D}_s \int^\infty_{-\infty} d z \,\, \vert \phi_1 (z) \vert^4 =
 \frac{3 u^{\rm 1 D}_s}{\sqrt{32 \pi} l_0}, \nonumber \\
 V = u^{\rm 1 D}_s \int^\infty_{-\infty} d z \,\, \vert \phi_0 (z) \phi_1 (z)
 \vert^2 = \frac{u^{\rm 1 D}_s}{\sqrt{8 \pi} l_0}, \nonumber
\end{align}
where $\phi_n (z) = h_n (z / l_0) \, \e^{-z^2 / 2 l_0^2} / \sqrt{2^n n !
\sqrt{\pi} l_0}$, $h_n (\xi)$ are Hermite polynomials \cite{gradshteyn-2014-1},
and $l_0 = 1 / \sqrt{m \omega}$.
The coefficient $u^{\rm 1 D}_s$ is a ``projected'' three-dimensional energy
scale $u_s$ that takes into account the transverse confinement:
\begin{displaymath}
 u^{\rm 1 D}_s = u_s \int d x \, d y \,\, \vert \phi_\perp (x, y) \vert^4,
\end{displaymath}
$\phi_\perp (x, y)$ is a transverse mode that we assume to be the same for both
longitudinal states $\phi_0 (z)$ and $\phi_1 (z)$.

\begin{figure}[t]
 \begin{center}
  \includegraphics[width = \columnwidth]{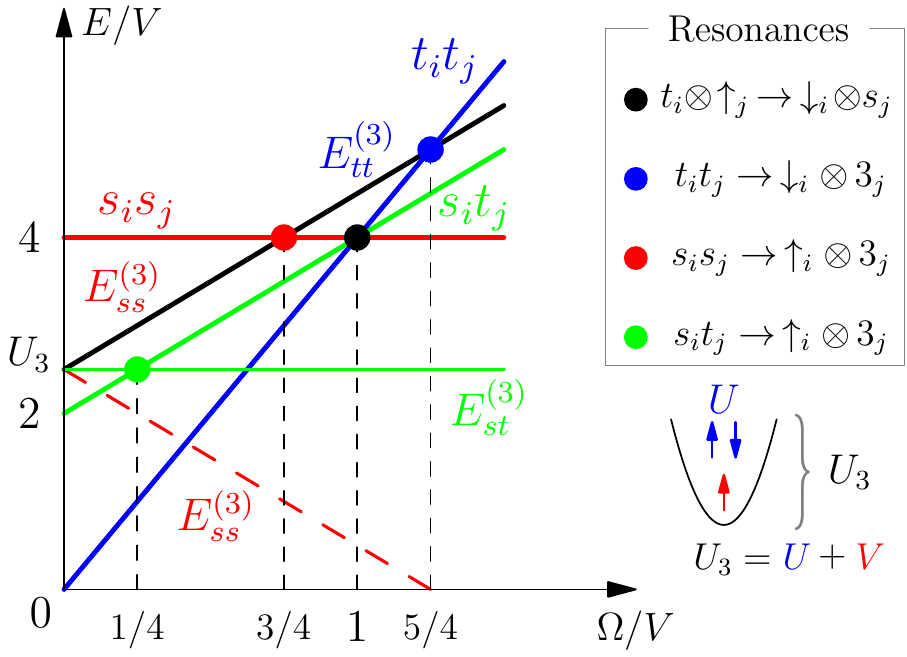}
 \end{center}
 \caption{
  Energies of the four-fermion system on two lattice sites as functions of the
  staggered field $\Omega$ (see text for notations).
  Red, blue and green lines correspond to $\vert s_i s_j \rangle$, $\vert t_i
  t_j \rangle$ and $\vert s_i t_j \rangle$ initial states, respectively (as
  indicated in the plot).
  Other lines of the same color are energies of the states $\vert
  \sigma_i\rangle \otimes \vert 3, \sigma^\prime \rangle_j$ obtained from the
  initial ones after single-fermion hopping.
  The black line includes two energy levels: $E^{(3)}_{t t}$ (blue) and
  $E^{(3)}_{s s}$ (red).
  The dashed red line corresponds to $E^{(3)}_{s s}$ which can not become
  resonant with other (red-colored) states at $\Omega > 0$.
  The circles mark different resonances: the singlet-triplet crossing in Fig.
  1(c) is identified by the black circle; other colors denote situations when
  one of the states involves a doubly occupied excited band.
  These states are off-resonant and can be ignored as long as the system is
  not too far detuned from the s-t resonance (i.e. $\Delta < V / 4$).
 }
 \label{fig:sfig2}
\end{figure}

There are two degenerate three-particle states: $\vert 3, \sigma \rangle =
\cd_{1 \sigma} \cd_{2 \ua} \cd_{2 \da} \vert 0 \rangle$ with an energy $U_3 = U
+ V = 5 u^{\rm 1 D}_s / \sqrt{8 \pi} l_0$.
We would like these states to be off-resonant, so that no hopping process could
connect them to any state within the s-t subspace.
Such processes can be computed by applying the excited-band hopping
$\cd_{2, j \sigma} c_{2, i \sigma}$ ($i$ and $j$ are lattice sites) to each of
the two neighboring two-particle states $\vert s_i s_j \rangle$, $\vert s_i t_j
\rangle$ and $\vert t_i t_j \rangle$.
We will assume that in the original (lab) frame $\Omega_i = \Omega$, $\Omega_j
= -\Omega$, and therefore $\vert t_i \rangle = \cd_{2, i \da} \vert \da_i
\rangle$, $\vert t_j \rangle = \cd_{2, j \ua} \vert \ua_j \rangle$.
The resulting target states and their energies are: $\vert s_i s_j \rangle \to
\vert \da_i \rangle \otimes \vert 3, \ua \rangle_j + \vert \ua_i \rangle
\otimes \vert 3, \da \rangle_j$ [$E^{(3)}_{s s} = \pm 2 \Omega + U_3$,
respectively], $\vert s_i t_j \rangle \to \vert \ua_i \rangle \otimes \vert 3,
\ua \rangle_j$ [$E^{(3)}_{s t} = U_3$], $\vert t_i t_j \rangle \to \vert \da_i
\rangle \otimes \vert 3, \ua \rangle_j$ [$E^{(3)}_{t t} = 2 \Omega + U_3$].
In Fig. \ref{fig:sfig2} we compare $E^{(3)}$ with the sum of singlet and
triplet energies.
It follows that three-particle states are off-resonant if the detuning from the
s-t resonance $\Omega = V$ does not exceed $V / 4$.

\begin{figure*}[t]
 \begin{center}
  \includegraphics[width = 2.05 \columnwidth]{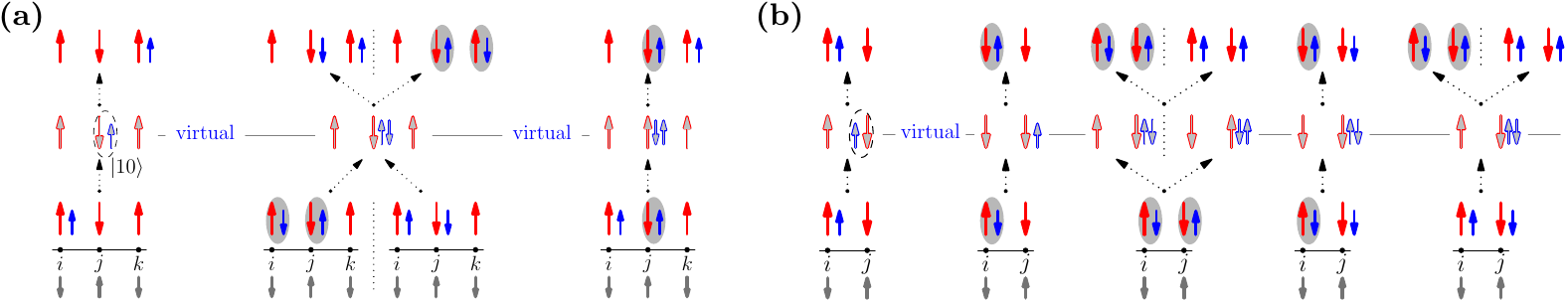}
 \end{center}
 \caption{
  Panel (a) Virtual processes leading to the next-NN correlated hopping in Eq.
  \eqref{eq:dirac-2nd-order-corrections}.
  Red (blue) arrows correspond to spins of local (itinerant) atoms.
  Virtual states are indicated by empty arrows.
  Thick gray arrows denote the artificial magnetic field $\Omega$.
  Filled gray ellipses are spin-singlet [$\vert 0 0 \rangle$] states, while the
  dashed ellipse denotes a $\vert 1  0 \rangle$ triplet.
  An itinerant atom is transferred from site $i$ to $k$ through an intermediate
  site $j$, whose state may change in the process.
  (b) Fluctuations that give rise to the density and exchange interactions in
  Eq. \eqref{eq:dirac-2nd-order-corrections}.
  Notations are the same as in panel (a).
 }
 \label{fig:sfig3}
\end{figure*}

\subsection{Corrections to the effective model in Eq. (4)}

The effective low-energy Hamiltonian $H_d$ in Eq. (4) approximates the full
model (5) to a leading (first) order in $J_0 / \Omega$ (or $J_0 / V$).
Here we compute second-order corrections $\sim J_0^2 / \Omega$ that involve
off-resonant states outside of the s-t subspace.
We will assume that $J_0$ and $\Delta$ are small compared to $V$ and $\Omega$,
and hence neglect the difference $V - \Omega$ in the energy denominators.
Our analysis is generic and valid in any space dimension.

There are two types of second-order corrections: next-NN (NNN) correlated
hopping, and density and exchange-like interactions.
The NNN hopping involves two links (three sites) and is conditional on the
state at an intermediate site [see Fig. \ref{fig:sfig3}(a)].
On the other hand, density and exchange-like interactions involve only pairs of
sites (single link), as show in Fig. \ref{fig:sfig3}(b).

We start by computing the action of a single kinetic-energy link $l_{i j} =
-J_0 \sum_n (\cd_{i n} c_{j n} + {\rm h.c.})$ on two types of initial states:
$\vert \psi^1_{\rm in} \rangle = \dd_{j \beta} \vac$ and $\vert \psi^2_{\rm in}
\rangle = \dd_{i \alpha} \dd_{j \beta} \vac$ with $\alpha$ and $\beta$ being
either $s$ or $t$.
In practice, this is easier to accomplish in the staggered frame (2) when
$l_{i j} = -J_0 \sigma^x_{n m} (a^\dag_{i n} a_{j m} + {\rm h.c.})$.
Energies of these states are $E^1_{\rm in} = \bigl( \frac{3}{2} \Omega \bigr)_j
+ \Omega_i = \frac{5}{2} \Omega$ and $E^2_{\rm in} = 2 \times \frac{3}{2}
\Omega = 3 \Omega$.
In the intermediate states $l_{i j} \vert \psi^{1, 2}_{\rm in} \rangle$, we
keep only those components, orthogonal to the s-t manifold in Fig. 1(e) (marked
by a red circle).
We have:
\begin{displaymath}
 l_{i j} \vert \psi^1_{\rm in} \rangle \to \frac{J_0}{2} \delta_{\beta s}
 \bigl[ \underbracket[0.1pt]{
  \vert 0 0 \rangle_i \vert \ua_j \rangle
 }_{\color{blue} 3 \Omega / 2 - \Omega} + \underbracket[0.1pt]{
  \vert 1 0 \rangle_i \vert \ua_j \rangle
 }_{\color{blue} -\Omega / 2 - \Omega} \bigr] - \frac{J_0}{\sqrt{2}}
 \delta_{\beta t} \underbracket[0.1pt]{
  \vert 1 0 \rangle_i \vert \da_j \rangle
 }_{\color{blue} -\Omega / 2 + \Omega}
\end{displaymath}
and for the state $\vert \psi^2_{\rm in} \rangle$
\begin{widetext}
 \begin{align}
  l_{i j} \vert \psi^2_{\rm in} \rangle \to & \frac{J_0}{2} \delta_{\alpha s}
  \delta_{\beta s} \underbracket[0.1pt]{
   (a^\dag_{i \ua} a^\dag_{i \da} - a^\dag_{j \ua} a^\dag_{j \da}) \vert \ua_i
   \ua_j \rangle
  }_{\color{blue} U - 2 \Omega} + J_0 \biggl[ \delta_{\alpha t}
  \delta_{\beta t} - \frac{1}{2} \delta_{\alpha s} \delta_{\beta s} \biggr]
  \underbracket[0.1pt]{
   (a^\dag_{i \ua} a^\dag_{i \da} - a^\dag_{j \ua} a^\dag_{j \da}) \vert \da_i
   \da_j \rangle
  }_{\color{blue} U + 2 \Omega} - \nonumber \\
  & - \frac{J_0}{\sqrt{2}} \underbracket[0.1pt]{
   (a^\dag_{i \ua} a^\dag_{i \da} - a^\dag_{j \ua} a^\dag_{j \da}) \bigl[
   \delta_{\alpha s} \delta_{\beta t} \vert \ua_i \da_j \rangle +
   \delta_{\alpha t} \delta_{\beta t} \vert \da_i \ua_j \rangle \bigr]
  }_{\color{blue} U}, \nonumber
 \end{align}
\end{widetext}
where energies of the intermediate states are shown as blue under-scripts.
Applying the link operators to the right-hand sides of these expressions, we
arrive at a second-order correction to the effective model in Eq. (4)
\begin{widetext}
 \begin{align}
  H^{(2)}_{\rm ef} = & \frac{J_0^2}{4 \Omega} \sum_{\triangle (i j k)} (1 -
  n^d_i) \bigl( \dd_{k t} d_{j s} + {\rm h.c.} \bigr) + \frac{J_0^2}{4 \Omega}
  \sum_{\langle i j \rangle} \biggl[ (1 - n^d_i) n^d_{j t} + \frac{3}{4} (1 -
  n^d_i) n^d_{j s} + (i \leftrightarrow j) \biggr] +
  \label{eq:dirac-2nd-order-corrections} \\
  & + \frac{J_0^2}{\Omega - U} \sum_{\langle i j \rangle} \biggl[
  \frac{3 \Omega - U}{5 \Omega - U} n^d_{i s} n^d_{j s} + 2 n^d_{i t}
  n^d_{j t} + \frac{\Omega - U}{3 \Omega - U} \bigl( n^d_{i s} n^d_{j t} +
  n^d_{i t} n^d_{j s} \bigr) - \bigl( \dd_{i s} \dd_{j s} d_{j t} d_{i t} +
  \dd_{i t} \dd_{j t} d_{j s} d_{i s}\bigr) \biggr] - \nonumber \\
  & - \frac{J_0^2}{\Omega - U} \sum_{\triangle (i j k)} \biggl[ \dd_{k t}
  \dd_{i t} d_{i s} d_{j s} + \dd_{k s} \dd_{i s} d_{i t} d_{j t} - \frac{1}{2}
  \dd_{k s} n^d_{i s} d_{j s} - 2 \dd_{k t} n^d_{i t} d_{j t} -
  \frac{\Omega - U}{3 \Omega - U} \dd_{k t} n^d_{i s} d_{j t} +
  (i \leftrightarrow j) + {\rm h. c.} \biggr]. \nonumber
 \end{align}
\end{widetext}
The 1st [2nd and 3rd] line corresponds to $\vert \psi^1_{\rm in} \rangle$
[$\vert \psi^2_{\rm in} \rangle$], and $\sum_{\triangle (i j k)}$ denotes
summation over all triples of sites $i$, $j$ and $k$.
In 1D it corresponds to second-nearest neighbors, while in 2D -- to second- and
third-nearest neighbors.

\section{Wavepacket dynamics with non-interacting fermions in 1D}

In the main text we presented evolution of many-body wavepackets that can be
determined only numerically.
Here we illustrate hallmark properties of the model (4), such as emergence
of the transverse local magnetic polarization, by studying the case of
canonical (i.e. non-interacting) $d$-fermions when Eq. (4) is the complete
Hamiltonian of the system and the Schrodinger equation can be solved
analytically for any initial condition.

\begin{figure}[t]
 \begin{center}
  \includegraphics[width = 0.8 \columnwidth]{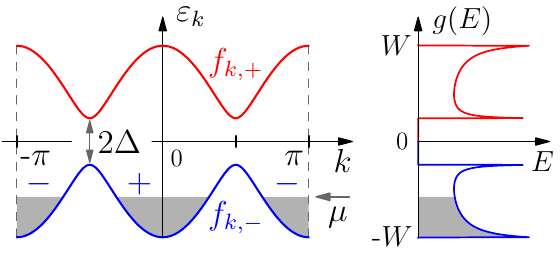}
 \end{center}
 \caption{
  Band structure of quasiparticles in Eq. \eqref{eq:quasiparticles} (left) and
  their density of states \eqref{eq:f-dos} (right).
  $W = \sqrt{(2 J)^2 + \Delta^2}$ was defined in Fig. 3.
  Gray shading indicates the filled Fermi sea; $\mu$ is the corresponding
  chemical potential.
 }
 \label{fig:sfig4}
\end{figure}

We can straightforwardly diagonalize the Hamiltonian (4) by rewriting it in the
momentum space:
\begin{displaymath}
 H = \sum_k \dd_k \bigl[ -2 J \cos k \, \sigma^x + \Delta \sigma^z \bigr] d_k.
\end{displaymath}
For a fixed momentum $k$, there are two quasiparticle states $f_{k, \pm}$ with
energies $\ve_{k, \pm} = \pm \rho_k$, $\rho_k = \sqrt{(2 J \cos k)^2 +
\Delta^2}$ shown in the left panel in Fig. \ref{fig:sfig4}.
The $d$-operators can be written as:
\begin{equation}
 \begin{pmatrix}
  d_{k s} \\ d_{k t}
 \end{pmatrix}
 =
 \begin{pmatrix}
  \cos \frac{\vartheta_k}{2} f_{k, +} - \sin \frac{\vartheta_k}{2} f_{k, -} \\
  \sin \frac{\vartheta_k}{2} f_{k, +} + \cos \frac{\vartheta_k}{2} f_{k, -}
 \end{pmatrix}
 =
 \begin{pmatrix}
  \sum_\lambda V^\lambda_k f_{k \lambda} \\
  \sum_\lambda U^\lambda_k f_{k \lambda}
 \end{pmatrix}
 \label{eq:quasiparticles}
\end{equation}
with ${\rm tg} \, \vartheta_k = -2 J \cos k / \Delta$ and $\lambda = \pm$.

\subsection{Single-fermion case}

Let us first consider a situation with only one fermion, whose wavefunction at
$t = 0$ is a fully polarized (triplet) Gaussian wavepacket centered at $x_i =
x_0$, with a width $\sigma$ and momentum $k_0$:
\begin{displaymath}
 \vert \psi_0 \rangle = A \sum_i \e^{-(x_i - x_0)^2 / 2 \sigma^2}
 \e^{\ii k_0 x_i} \dd_{i t} \vac.
\end{displaymath}
This state is normalized as $\langle \psi_0 \vert \psi_0 \rangle = 1$ ($A$ is
the normalization constant).
The time-dependent solution can be readily written as:
\begin{align}
 \vert \psi (t) \rangle = & \sum_k \beta (k) \bigl( \sin {\textstyle
 \frac{\vartheta_k}{2}} \e^{-\ii \ve_{k, +} t} \fd_{k, +} + \nonumber \\
 & + \cos {\textstyle \frac{\vartheta_k}{2}} \e^{-\ii \ve_{k, -} t} \fd_{k, -}
 \bigr) \vac, \nonumber
\end{align}
where $\beta (k)$ is a Fourier transform of the original packet
\begin{equation}
 \beta (k) = \frac{A}{\sqrt{N}} \sum_i \e^{-(x_i - x_0)^2 / 2 \sigma^2}
 \e^{\ii (k_0 - k) x_i}.
 \label{eq:wp-fourier}
\end{equation}
We are interested in the time-dependent local spin-resolved density
$n^d_{i \alpha} (t) = \langle \psi (t) \vert \dd_{i \alpha} d_{i \alpha} \vert
\psi (t)\rangle$ (no summation over $\alpha$) and transverse spin polarization
$T^x_i (t) = -\frac{1}{\sqrt{8}} \sigma^x_{\alpha \beta} \langle \psi (t) \vert
\dd_{i \alpha} d_{i \beta} \vert \psi (t) \rangle$.
Using the relation $\dd_{i \alpha} d_{i \beta} = \frac{1}{N} \sum_{p^\prime p}
\e^{\ii (p - p^\prime) x_i} \dd_{p^\prime \alpha} d_{p \beta}$, one can show
that $n^d_{i \alpha} (t) = \vert C_{i \alpha} (t) \vert^2$ and $T^x_i (t) =
-\frac{1}{\sqrt{8}} C^*_{i \alpha} (t) \sigma^x_{\alpha \beta} C_{i \beta} (t)$
with
\begin{align}
 \begin{pmatrix}
  C_{i s} \\ C_{i t}
 \end{pmatrix}
 \! = \! \frac{1}{\sqrt{N}} \sum_k \beta (k) \e^{\ii k x_i} \!
 \begin{pmatrix}
  \ii \frac{2 J \cos k}{\rho_k} \sin \rho_k t \\
  \cos \rho_k t + \ii \frac{\Delta}{\rho_k} \sin \rho_k t
 \end{pmatrix}
 \label{eq:1-particle-wavefunction}
\end{align}
$C_{i \alpha}$ is a single-particle real-space wavefunction.

Due to the presence of negative energies $\ve_{k, -}$, the initial
wavepacket splits into two counter-propagating, right and left moving parts
(mathematically this happens because the time dependence enters only via $\cos
\rho_k t$ and $\sin \rho_k t$).
Indeed, assuming that $\beta (k)$ is peaked near the initial momentum $k_0$
with a width $\Delta k$, we can approximately compute the above sums:
\begin{align}
 C_{i t} \sim & \int_{k_0 - \Delta k}^{k_0 + \Delta k} \!\! d k \,\, \beta (k)
 \, \e^{\ii k_0 x_i} \sum_{\lambda = \pm} \frac{\rho_k + \lambda \Delta}{2
 \rho_k} \, \e^{\ii \lambda \rho_k t} \approx \nonumber \\
 \approx & \int_{-\Delta k}^{\Delta k} \!\! d k \, \beta (k + k_0) \,
 \e^{\ii (k + k_0) x_i} \! \sum_{\lambda = \pm} \frac{\rho_{k_0} + \lambda
 \Delta}{2 \rho_{k_0}} \, \e^{\ii \lambda \rho_{k + k_0} t}. \nonumber
\end{align}
Introducing the group velocity $v_{k_0} = -\partial \rho_k / \partial k_0$, we
have: $C_{i t} \approx \sum_\lambda \bigl[ 1 + \frac{\lambda
\Delta}{\rho_{k_0}} \bigr] w_t (x_i - \lambda v_{k_0} t) = r_t (x_i - v_{k_0}
t) + l_t (x_i + v_{k_0} t)$, where $w_t (\xi)$ is defined by the initial state
and the two terms correspond to right and left movers (for $v_{k_0} > 0$).
A similar manipulation for $C_{i s}$ yields: $C_{i s} \approx
\frac{J}{\rho_{k_0}} \cos k_0 \sum_\lambda w_s (x_i - \lambda v_{k_0} t)$.
In general for $\Delta \neq 0$, $\vert l_t \vert < \vert r_t \vert$ (for the
same value of their arguments) and the two wavepackets are not symmetric.
However, exactly at the s-t resonance $\Delta = 0$, $l_t = r_t$, so left and
right movers are mirror images of each other.
For singlets, $r_s = -l_s$.
This phase difference leads to an opposite sign of transverse local
magnetization $\langle T^x_i (t) \rangle$ for left and right moving parts of
the distribution.

\subsection{Many-particle wavepackets}

The results obtained for a single $d$-fermion allow us to investigate wavepacket
dynamics with several particles.
Specifically, we focus on the five-fermion case considered in the main text
and derive a closed expression for the wavefunction and evolution of the total
density.

We assume that the initial wavefunction contains only triplets, and is a ground
state (GS) in a harmonic trap $V (x_i) = A \bigl( x_i - \frac{N}{2} \bigr)^2$
imposed on a chain with open boundary conditions ($x_i = i = 0 \ldots N - 1$):
\begin{align}
 \vert \psi_{N_d}^{\lbrace n \rbrace} & (t = 0) \rangle = \prod_\nu
 \dd_{n_\nu t} \vac = \sum_{\lbrace i \rbrace} \e^{\ii k_0 \sum_\nu x_{i_\nu}}
 \times \nonumber \\
 \times & \prod_\nu \phi_{n_\nu} (x_{i_\nu}) \dd_{i_\nu t} \vac =
 \sum_{\lbrace k \rbrace} \prod_\nu \beta_{n_\nu} (k_\nu) \dd_{k_\nu t} \vac,
 \nonumber
\end{align}
where $\phi_n (x_i)$, $n = 0, 1, \ldots$ are single-particle eigenfunctions in
the trap [$\phi_n (-1) = \phi_n (N) = 0$] and $\beta_n (k)$ is the Fourier
transform of $\phi_n (x) \e^{\ii k_0 x}$ defined in Eq. \eqref{eq:wp-fourier}
for a single mode.
The total number of fermions is $N_d$, $\nu = 1 \ldots N_d$ and $k_0$ is the
center of mass momentum.
Finally, $\lbrace a \rbrace = a_1 \ldots a_{N_d}$ with $a = i$, $n$, etc.

At time $t = 0$, the trap is removed and the wavepacket starts to propagate.
The time-dependent wavefunction $\vert \psi^{\lbrace n \rbrace}_{N_d} (t)
\rangle$ can be written using $f$-quasiparticles \eqref{eq:quasiparticles}:
\begin{displaymath}
 \vert \psi^{\lbrace n \rbrace}_{N_d} (t) \rangle =
 \sum_{\lbrace k \lambda \rbrace} \prod_\nu \beta_{n_\nu} (k_\nu)
 U^{\lambda_\nu}_{k_\nu} \e^{-\ii \varepsilon_{k_\nu \lambda_\nu} t}
 \fd_{k_\nu \lambda_\nu} \vac.
\end{displaymath}
A straightforward calculation yields the position- and time-dependent total
density:
\begin{displaymath}
 \langle n^d_i (t) \rangle = \langle \psi^{\lbrace n \rbrace}_{N_d} (t) \vert
 \dd_{i \alpha} d_\alpha \vert \psi^{\lbrace n \rbrace}_{N_d} (t) \rangle =
 \sum_{\nu \alpha} \vert C^{n_\nu}_{i \alpha} (t) \vert ^2,
\end{displaymath}
where $C^{n_\nu}_{i \alpha} (t)$ is a multi-mode generalization of
$C_{i \alpha}$ in Eq. \eqref{eq:1-particle-wavefunction}, with $\beta_k \to
\beta_{n_\nu} (k)$.
$\langle n^d_{i_0} (t) \rangle$ is plotted in the inset of Fig. 2.

\section{Drude weight of non-interacting $d$-fermions}

Although we are interested in metallic properties of the strongly-correlated
model (4), it is nevertheless instructive to study its non-interacting limit
(relevant for the low-density regime), i.e. treat $d_{i \alpha}$ as
unconstrained, canonical fermions, and compute properties of this model such as
the GS energy and Drude weight ${\cal D}$ (as functions of the chemical
potential $\mu$), and the density of states.

The $d$-fermions on a ring pierced by a flux are described by a Hamiltonian (we
call it $H_0$ instead of $H_d$ to emphasize lack of correlations):
\begin{align}
 H_0 (\phi) = &\sum_i \bigl[ -J \sigma^x_{\alpha \beta} \bigl( \dd_{i \alpha}
 \e^{\ii \phi} d_{i + 1, \beta} + {\rm h.c.} \bigr) + \nonumber \\
+ \Delta & \sigma^z_{\alpha \beta} \dd_{i \alpha} d_{i \beta} \bigr] = \sum_k
 \bigl[ \epsilon_k (\phi) \sigma^x + \Delta \sigma^z \bigr]_{\alpha \beta}
 \dd_{k \alpha} d_{k \beta}, \nonumber
\end{align}
with $\epsilon_k (\phi) = -2 J \cos (k + \phi)$.
Its GS energy ${\cal E}_0 (\phi)$ is
\begin{displaymath}
 \frac{{\cal E}_0 (\phi)}{N} = \frac{1}{N} \sum_k \bigl[ \epsilon_k (\phi) \sin
 \vartheta_k + \Delta \cos \vartheta_k \bigr] (n^f_{k, +} - n^f_{k, -}),
\end{displaymath}
where $n^f_{k \lambda} = \theta (\mu - \varepsilon_{k \lambda})$ is the
zero-temperature Fermi function.
The momentum integral in this expression is trivially computed and we have
\begin{align}
 & \frac{{\cal E}_0 (\phi = 0)}{N} = -\frac{2 W}{\pi} \,\, E \biggl( \arcsin
 \frac{\sqrt{W^2 - \mu^2}}{2 J}, \frac{2 J}{W} \biggr), \nonumber \\
 & {\cal D} = \frac{1}{N} \frac{\partial^2 {\cal E}_0}{\partial \phi^2} \biggl
 \vert_{\phi = 0} = \frac{4 J}{\pi} \,\, E \biggl( \arcsin
 \frac{\sqrt{W^2 - \mu^2}}{W}, \frac{W}{2 J} \biggr). \nonumber
\end{align}
Here $E (a, m) = \int_0^{\sin a} d x \, \sqrt{1 - m^2 x^2} / \sqrt{1 - x^2}$
is an incomplete elliptical integral of the second kind
\cite{gradshteyn-2014-1}, and $W = \sqrt{(2 J)^2 + \Delta^2}$.
These functions are shown in the inset of Fig. 3(a).

Finally, the spin-resolved density of states of the $f$-quasiparticles, plotted
in the right panel of Fig. \ref{fig:sfig4}, is
\begin{align}
 g (E) = & \frac{1}{N} \sum_{k \lambda} \delta (E - \varepsilon_{k \lambda}) =
 \nonumber \\
 & = \frac{2 \vert E \vert \, \theta (\vert E \vert - \Delta) \, \theta (W -
 \vert E \vert)}{\pi \sqrt{(E^2 - \Delta^2) (W^2 - E^2)}}.
 \label{eq:f-dos}
\end{align}

\begin{figure}[t]
 \begin{center}
  \includegraphics[width = \columnwidth]{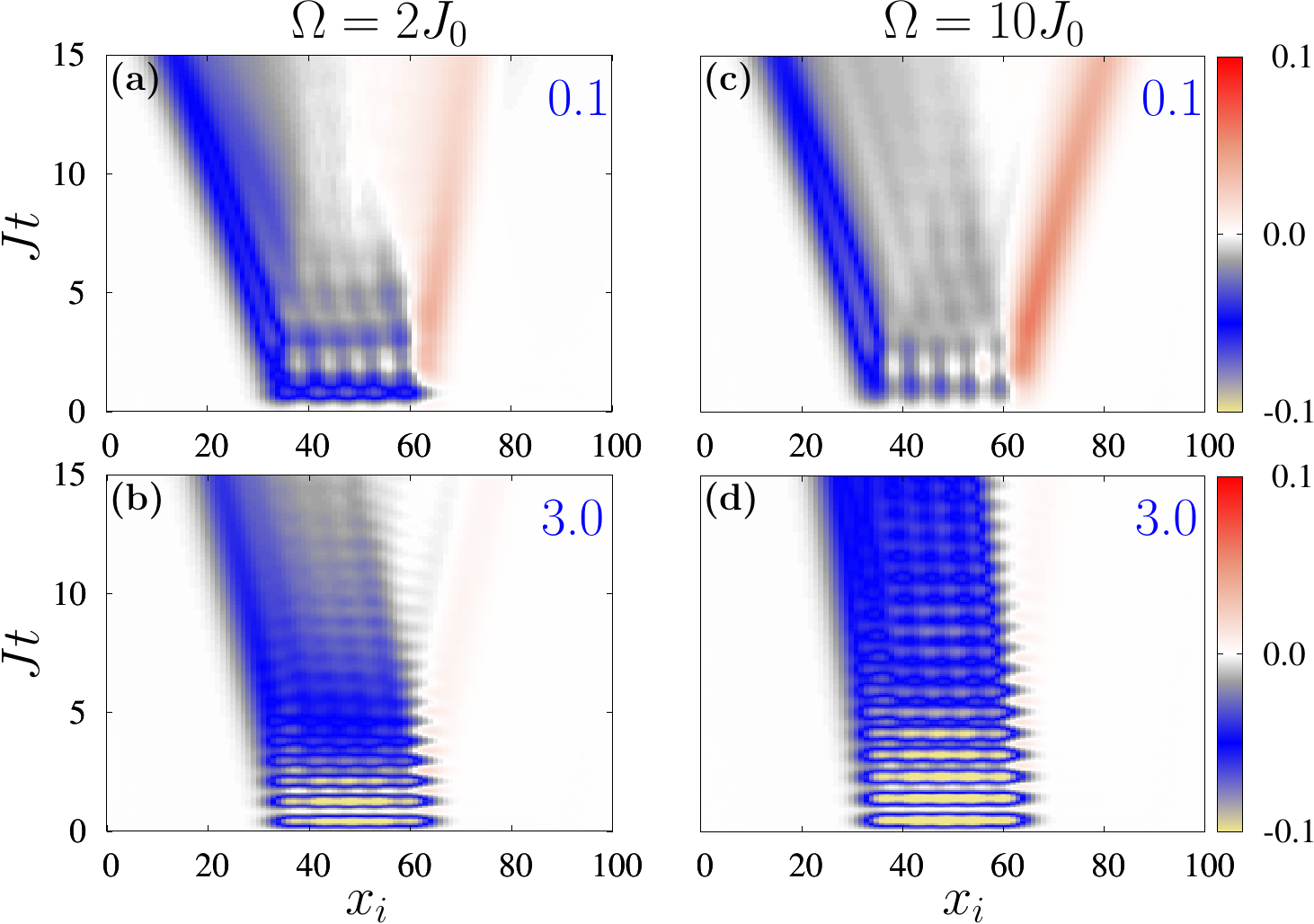}
 \end{center}
 \caption{
  Dynamics of many-body wavepackets with the same parameters as in Fig. 2, but
  computed within the full AEA Hamiltonian (5). Shown is the evolution of the
  transverse spin polarization $\langle T_i^x \rangle$.
  Panels {\bf (a)} and {\bf (b)} [{\bf (c)} and {\bf (d)}] correspond to Rabi
  frequency $\Omega = 2 J_0$ ($\Omega = 10 J_0$).
  The detuning $\Delta$ is chosen to match Fig. 2: $\Delta = 0.1 J$ [(a) and
  (c)], and $\Delta = 3 J$ [(b) and (d)] with $J = J_0 / \sqrt{2}$.
 }
 \label{fig:sfig5}
\end{figure}

\section{Details of numerical calculations}

To obtain results in the main text we performed unbiased DMRG calculations for
the effective model (4).
In our DMRG computations quantum states are represented by matrix product
states (MPS) \cite{schollwock-2011-1} with a particular ``bond-dimension'' $D$.
In the limit of large $D$, this state representation becomes exact.

To compute GS properties we write the Hamiltonian as a matrix product operator
(MPO) and apply a variational GS search that uses updates on neighboring sites
simultaneously \cite{schollwock-2011-1}.
Periodic boundary conditions are implemented via long-range hopping terms in
the MPO.
Since those terms significantly increase correlations over the length of the
chain, periodic boundary conditions make the calculations significantly more
challenging and require larger values of $D$.
In practice, the three-dimensional local Hilbert space of the effective model
(4) allows us to scan parameter regime for a system of $N = 40$ sites within
reasonable CPU times.
The local Hilbert space dimension of the full model (5) is $8$, and requires
significantly longer CPU times.
For the GS phase diagram in Fig. 3 we use up to $D = 265$ and find that the
results are typically well converged for $D = 128$.
The convergence is reached when relative change in energy is less than
$10^{-8}$, with the corresponding maximum truncated weights for two-site
updates are $\sim 10^{-7}$.

For the time-evolution calculations we use a time-dependent block decimation
algorithm \cite{vidal-2004-1,white-2004-1,daley-2004-1}, which approximates
an application of the time-evolution operator by a 4th order Trotter
decomposition.
For the many-body wave-packet dynamics in Fig. 2, the results are well
converged for bond dimension $D = 128$.

\begin{figure}[t]
 \begin{center}
  \includegraphics[width = \columnwidth]{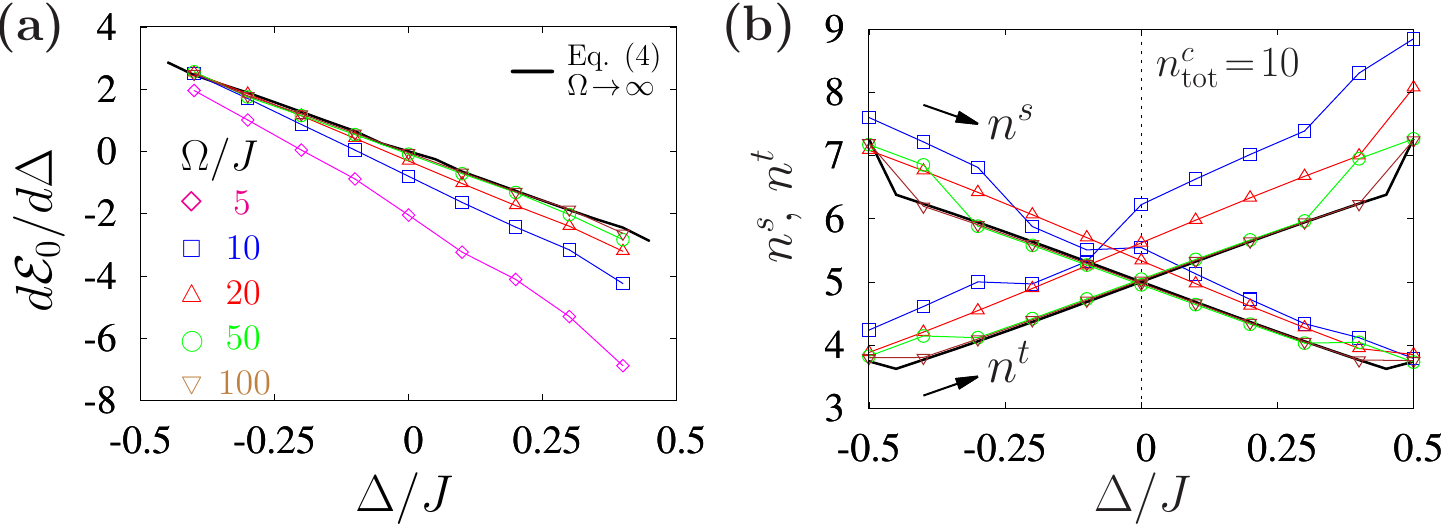}
 \end{center}
 \caption{
  GS observables within the full Hamiltonian (5) and the effective model (4).
  The system size is $N = 30$ with box boundary conditions, and the chemical
  potential is $\mu = -J$.
  {\bf (a)} Derivative of the GS energy w.r.t. $\Delta$ for increasing
  $\Omega / J$: 5 (magenta diamonds), 10 (blue squares), 20 (red triangles), 50
  (green circles), and 100 (brown inverted triangles).
  Solid black line is computed withing the effective model (4).
  {\bf (b)} The number of effective singlets ($n^s$) and triplets ($n^t$).
  The total number of mobile atoms is $n^c_{\rm tot} = \sum_{i n} n^c_{i 2 n} =
  10$.
  Other notations are the same as in panel (a).
 }
 \label{fig:sfig6}
\end{figure}

\subsection{Validity of effective model in Eq. (4)}

In the main text, numerical calculations focused on the effective Dirac-like
model (4).
Here we verify its validity for the GS calculations in Fig. 3.

We compare GS observables obtained within DMRG for the effective model (4) and
the full AEA model (5), which is expected to reduce to Eq. (4) in the limit of
large $\Omega$ and small $\Delta \lesssim J$, in a system of $N = 30$ sites
with box boundary conditions.
To converge to a sector with the same number of particles we add a chemical
potential term $\bigl[ \frac{1}{2} (V + \Delta) + \mu \bigr] \sum_i
(n^c_{i 2 \ua} + n^c_{i 2 \da})$ to Eq. (5) and compute GS properties for
$\mu = -J$ and $-\frac{1}{2} \leqslant \Delta / J \leqslant \frac{1}{2}$.

In Fig. \ref{fig:sfig6} we present a direct comparison of the GS energy
derivative $d {\cal E}_0 / d \Delta$ [panel (a)], and the number of local
singlets $n^s = \sum_i \vert s \rangle_i \langle s \vert_i$ with $\vert s
\rangle_i = \frac{1}{\sqrt{2}} \bigl[ \ad_{i \ua} \vert \da \rangle_i -
\ad_{i \da} \vert \ua \rangle_i \bigr]$ and triplets $n^t = \sum_i \vert t
\rangle_i \langle t \vert_i$ with $\vert t \rangle_i = \ad_{i \da} \vert \da
\rangle_i$ [cf. Eq.  (3)] in the system [panel (b)].
Clearly in the limit of large $\Omega$ the results converge to the ones
obtained within the effective model (4).
Moreover, Fig. \ref{fig:sfig6}(b) demonstrates that for increasing $\Omega$ the
population in states outside the s-t manifold vanishes.
Indeed, for the parameters of the figure the total number of particles is
fixed, $n^a = 10$, and for large $\Omega$ all of them belong to the s-t
subspace.
All these results confirm that the physics of the problem is captured by the
low-energy model (4) in the strong-coupling regime $V \sim \Omega \gg J_0$.

\subsection{Wave-packet dynamics in the full AEA model}

For the wave-packet dynamics calculations in Fig. 2 we used the effective model
(4).
Here we demonstrate numerically that characteristic features of this dynamics
are also present in the full AEA Hamiltonian (5), and can be observed in
realistic experiments.

We use the same initial state as the one described in the main text, i.e. a
wavepacket that consists of five triplets, and is produced by introducing a
harmonic trapping potential and computing the ground state of the system.
At time $t=0$, the trap is removed and the wavepacket is accelerated by
applying a phase-gradient operator.
The interband exchange interaction, as well as the Hubbard repulsion are set by
the detuning and Rabi frequency: $V = \Delta + \Omega$ and $U = 3 V / 2$,
respectively (see previous section).
Fig. \ref{fig:sfig5} shows the transverse spin polarization $\langle T_i^x
\rangle$ and demonstrates that for large $\Omega = 10 J_0$ the exact dynamics
of the Dirac-like Hamiltonian (4) of Fig. 2 is reproduced.
For small $\Omega$ the effective model starts to break down, but remarkably
even for $\Omega = 2 J_0$, the full time evolution still features the
characteristic splitting into two counter-propagating parts with opposite
$\langle T_i^x \rangle$ polarization for small $\Delta < J, J_0$.

\bibliographystyle{apsrev4-1}
\bibliography{lit}